\documentclass[11pt]{article}
\usepackage{authblk}
\usepackage[margin=1.1in]{geometry}
\usepackage{amsmath,amssymb,amsfonts,graphicx,float}
\usepackage{bm}
\usepackage{booktabs}
\usepackage{xcolor}  
\usepackage[numbers]{natbib}
\usepackage[colorlinks=true,linkcolor=blue,citecolor=blue]{hyperref}
\usepackage{subcaption}
\usepackage{microtype}

\newcommand{\doi}[1]{\href{https://doi.org/#1}{\nolinkurl{#1}}}

\newcommand{\dno}{G(\eta)}            
\newcommand{\fno}{\mathcal{S}_\theta} 
\newcommand{\R}{\mathbb{R}}
\newcommand{\T}{\mathbb{T}}
\newcommand{\eps}{\varepsilon}

\title{Learning the solution operator of standing waves \protect\\ with a Fourier neural operator}

\author[1]{Changhong Mou\thanks{E-mail:~changhong.mou@usu.edu}}
\author[2]{Savana Ammons\thanks{E-mail:~savanaa2@illinois.edu}}
\author[3]{Di Qi\thanks{Email:~qidi@purdue.edu}}
\author[4]{Vera Mikyoung Hur\thanks{E-mail:~verahur@illinois.edu}}

\affil[1]{Department of Mathematics and Statistics, Utah State University, \protect\\ Logan, UT 84322, USA}
\affil[2]{Mobius Logic Inc., Tysons, VA 22102, USA}
\affil[3]{Department of Mathematics, Purdue University, West Lafayette, IN 47907, USA}
\affil[4]{Department of Mathematics, University of Illinois Urbana-Champaign,\protect\\ Urbana, IL 61801, USA}

\date{\today}

\begin{document}
\maketitle

\begin{abstract}
The standing wave problem concerns the space-time periodic motion at the free surface of a two-dimensional, infinitely deep, irrotational flow of an incompressible inviscid fluid under gravity. Although spectrally accurate computations have revealed a rich array of nonlinear wave phenomena, repeatedly evaluating their evolution over large parameter spaces remains challenging, particularly when resolving fine-scale spatial oscillations. We employ a Fourier neural operator (FNO) architecture to learn the solution operator along a numerically computed one-parameter family of symmetric standing waves. Given a wave profile at an instant of rest and a normalized observation time, the learned operator predicts the fluid surface, the velocity potential and the dipole density at the surface at that time. In-distribution tests yield uniformly small relative errors within the training range. Two out-of-distribution experiments characterize extrapolation toward contrasting physical regimes. Toward the free-fall limit, errors concentrate near the sharpening crest, consistent with the spectral truncation in the FNO architecture. Toward the linear limit, absolute errors remain small whereas relative errors increase as the amplitude vanishes. A residual of the governing equations, evaluated independently of reference solutions, complements these tests and provides a diagnostic for exploratory predictions beyond the available data. 
\end{abstract}

\newpage

\section{Introduction}\label{sec:intro}

The classical standing wave problem concerns the space-time periodic motion of the free surface bounding a two-dimensional, infinitely deep, irrotational flow of an incompressible inviscid fluid under gravity. Since the pioneering work of Boussinesq \cite{Boussinesq1877} and Rayleigh \cite{rayleigh1915}, analytical, computational, and experimental studies have explored nonlinear wave phenomena arising in this canonical interfacial fluids problem. Here we consider a numerically computed one-parameter family of symmetric standing waves, spanning small to large amplitudes. For large amplitudes, the acceleration of a fluid particle at the crest approaches gravitational acceleration at an instant of rest. 

The limiting configuration of standing waves has motivated extensive investigation. Penney and Price \cite{penney1952} conjectured that a limiting wave exhibits a $90^\circ$ angle at the crest at rest, a hypothesis later supported by Taylor's experiments \cite{taylor1953}. Subsequent numerical studies  \cite{mercer1992,bryant1994,schultz1998,smith1999,okamura2003} developed increasingly detailed accounts of large amplitude waves, but they also revealed the difficulty of inferring a limiting profile from finite resolution computations. More recently, Wilkening \cite{wilkening2011breakdown} (see also \cite{wilkening2012overdetermined}) resolved fine-scale oscillations near the sharpening crests of `almost' limiting waves, demonstrating the breakdown of a self-similar approach to an angle. The rigorous mathematical treatment of Iooss, Plotnikov, and Toland \cite{iooss2005standing}, for small amplitudes, requires the Nash--Moser implicit function theorem for small divisor problems. Together, these results underscore how challenging the standing wave problem is for analysis and computation. 

The spectrally accurate computations of Wilkening and Yu \cite{wilkening2012overdetermined} provide the foundation for the present study, where a rest state and a temporal period are determined by solving a nonlinear boundary value problem, subject to periodicity and symmetry conditions. This requires high-order time integrations of high-dimensional systems of equations. The overdetermined shooting method in \cite{wilkening2012overdetermined} lowers computational cost through efficient linearized solves, economical Jacobian updates, and parallel computing. 
Nonetheless, repeatedly evaluating the evolution over large parameter spaces remains costly, particularly when fine spatial scales must be resolved. Furthermore, each numerical run yields a solution only at one parameter value, and the endpoint of the computed solution family is where the computation becomes prohibitive, rather than an established endpoint of the solution set. A learned representation may offer a complementary means of exploiting these numerical solutions, where the primary computational cost is incurred during offline data generation and training, while enabling online evaluation across a range of parameters at a fraction of the cost. This motivates two questions: How accurate is a surrogate within the training range, and how effectively does it generalize outside the range?

Prior research has demonstrated that standing waves admit such a surrogate. For instance, Williams \emph{et al.} \cite{williams2012low} established low dimensionality through proper orthogonal decomposition and constructed reduced-order models from a sparse set of modes. Their work serves as a direct precursor to the present study.

Machine learning methods have successfully addressed a variety of fluid mechanics problems in recent years \cite{brunton2020machine,vinuesa2022enhancing}. 
Neural operator learning, using architectures such as the deep operator network (DeepONet) \cite{lu2021deeponet}, the Fourier neural operator (FNO) \cite{li2021fourier}, and the general neural operator framework \cite{kovachki2023neural}, seeks to approximate solution operators of governing equations rather than their solutions, independently of the grid of the data.  
The FNO is particularly well-suited to the standing wave problem for several reasons. The spatial domain is periodic and the data are sampled on a uniform grid. Linear waves consist of the fundamental harmonic while nonlinear waves contain higher harmonics. Additionally, the Dirichlet-Neumann operator, in the formulation of the water wave problem in surface variables \cite{zakharov1968stability,craig1993numerical}, becomes a Fourier multiplier in the flat-surface limit, consistent with the spectral convolution underlying the FNO architecture. 

Furthermore, the standing wave problem provides an excellent example for assessing operator learning skill through interpolation and extrapolation errors. We remark that operator learning methods are commonly assessed primarily by interpolation errors \cite{li2021fourier, kovachki2023neural}. The dimensionless crest acceleration $\eps$ (see \eqref{eq:eps}) decreases monotonically along the dataset even where the wave amplitude or height has a turning point, serving as a parameter for numerical experiments. 
Toward the free-fall limit as $\eps\to 0$, the Fourier spectrum of the surface elevation decays slowly, making the reference solutions expensive to compute. Toward the linear limit as $\eps\to 1$, on the other hand, absolute errors remain small but relative errors increase as the amplitude vanishes. (This is a regular zero-amplitude limit but a singular limit with respect to a relative error measure.) A surrogate can therefore be driven toward each limit from within the training range, and its performance in these contrasting regimes can be examined individually. 
Additionally, the governing equations yield a residual that can be evaluated from the predictions, offering a complementary diagnostic even when ground-truth solutions are unavailable.

The present study makes three primary contributions to data-driven modeling of the standing wave problem. First, we introduce an FNO architecture to learn the solution operator from high-resolution reference solutions. Given a wave profile at an instant of rest and a normalized observation time, the learned operator predicts the fluid surface, the velocity potential and dipole density at the surface at that time. Evaluated on in-distribution (ID) test sets, the operator achieves relative $L^2$ errors below $10^{-3}$ within the training range. 
Second, we design two out-of-distribution (OOD) experiments, each withholding a contiguous block at one end of the training range. These experiments reveal different sources of error, the broadening Fourier spectrum toward the free-fall limit versus increasing relative errors toward the linear limit, which calls for different remedies. 
Third, we develop a diagnostic framework by combining errors that are measured against reference solutions with a residual that can be evaluated independently of reference solutions. In the OOD experiments, where reference solutions are available, we compare the residual with the prediction errors. For exploratory predictions beyond the available data, we assess candidate evolutions using the residual together with the responses of the models trained on different portions of the  dataset.

Our broader motivation is to explore strongly nonlinear waves using data-driven methods that provide a complementary use of conventional and learned approaches. Standing waves serve as an illustrative example. The low-cost learned operator, together with its diagnostics, provides an accurate representation within the sampled solution family while helping identify candidate evolutions for subsequent numerical correction.

The remainder of the manuscript is organized as follows. Section \ref{sec:problem} outlines the governing equations, the data, and their symmetries, introduces the downward particle acceleration at the crest at rest as a parameter, and summarizes the reference computations. Section \ref{sec:fno} formulates the operator learning problem and describes its FNO implementation. Section \ref{sec:results} presents the training and testing datasets, the training procedure, and the evaluation criteria, and discusses ID and OOD test results. Section \ref{sec:epstry} assesses the model's capacity for exploratory predictions beyond the available data. Section \ref{sec:conclusions} discusses the conclusions and future directions.


\section{Standing wave formulation and reference data}\label{sec:problem}

\subsection{Governing equations and stored fields}

We consider a two-dimensional, infinitely deep, irrotational flow of an incompressible inviscid fluid, under gravity, and the wave motion at its free surface. We take the Cartesian coordinates where waves propagate primarily in the $x$-direction and gravity acts in the negative $y$-direction. Suppose that the fluid occupies the region 
\[
\{(x,y)\in \mathbb{R}^2: y<\eta(x,t)\},\qquad t\in\mathbb{R},
\]
where $\eta(x,t)$ denotes the displacement of the fluid surface from a flat one, the first of the three fields stored in the dataset. 
The velocity potential $\Phi(x,y,t)$ satisfies
\begin{equation}\label{eq:Phi}
  \Delta\Phi=0 \quad \text{in $y<\eta(x,t)$}\quad\text{and}\quad  
  \nabla\Phi\to0 \quad \text{as $y\to-\infty$}.
\end{equation}
We normalize $\Phi$ and $\eta$ to be $2\pi$-periodic in the $x$-direction. The kinematic and dynamic free-surface conditions are
\begin{equation}\label{eq:BC}
\left.\begin{aligned}
&\eta_t=\Phi_y - \eta_x\Phi_x \\
&\Phi_t + \frac12|\nabla\Phi|^2 + g\eta = 0
\end{aligned}\right\}\qquad \text{on $y=\eta(x,t)$},
\end{equation}
where $g>0$ denotes gravitational acceleration. We choose the additive function of time in $\Phi$ so that the Bernoulli constant vanishes for all time. 

Let
\begin{equation*}
  \varphi(x,t)=\Phi(x,\eta(x,t),t)
\end{equation*}
denote the trace of the velocity potential on the fluid surface, the second stored field. The Dirichlet--Neumann operator $\dno$ maps $\varphi$ to the (scaled) outward normal derivative of its harmonic extension into the fluid. That is, 
\begin{equation*}
  \dno\varphi=\sqrt{1+\eta_x^2}[\partial_{\boldsymbol{n}}\Phi]_{y=\eta}
  =[\Phi_y-\eta_x\Phi_x]_{y=\eta},
\qquad \boldsymbol{n}=\frac{(-\eta_x,1)}{\sqrt{1+\eta_x^2}}.
\end{equation*}
Consequently, \eqref{eq:Phi} and \eqref{eq:BC} close in surface variables:
\begin{align}
  &\eta_t   = \dno\varphi, \label{eq:kin}\\
  &\varphi_t = -g\eta - \frac12\varphi_x^{2}
      + \frac{(\dno\varphi + \eta_x\varphi_x)^{2}}
             {2(1+\eta_x^{2})}. \label{eq:dyn}
\end{align}

Evaluating $\dno\varphi$ requires a harmonic extension of $\varphi$ into the fluid at each instant. The boundary integral method in \cite{wilkening2012overdetermined} represents this as a double layer potential with dipole density $\mu(x,t)$ on the fluid surface, the third stored field. More precisely, $\mu$ solves a second-kind Fredholm integral equation of the form
\begin{equation}
  \frac12\mu + K(\eta)\mu = \varphi,
  \label{eq:mu}
\end{equation}
where $K(\eta)$ is a boundary integral operator built from the periodic Green's function of the Laplacian in the fluid. See \cite{wilkening2012overdetermined} for its formula. Given $\mu$ in the dataset, $\dno\varphi$ can be evaluated to spectral accuracy. 

To recapitulate, the dataset stores three surface fields: $\eta$ and $\varphi$ are the dynamical variables of \eqref{eq:kin}-\eqref{eq:dyn}, and $\mu$ is the auxiliary variable from which the nonlocal operator is computed. These fields form the three outputs of our learned operator.

\subsection{Standing wave symmetries and data representation}\label{sec:symmetries}

\begin{figure}[ht]
  \centering
  \includegraphics[width=0.8\textwidth]{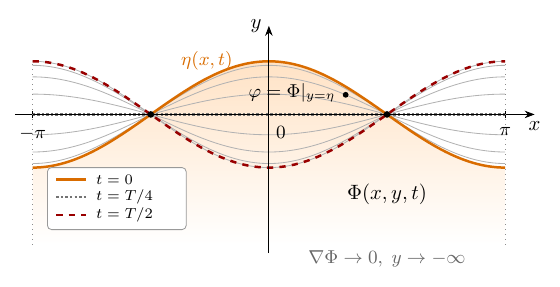}
\caption{
Schematic of a space-time periodic standing wave on $x\in[-\pi,\pi]$, using the dataset conventions. The fluid occupies $y<\eta(x,t)$, where $\Phi$ is harmonic, $\Phi=\varphi$ on $y=\eta(x,t)$, and $\nabla\Phi\to0$ as $y\to-\infty$. The solid orange curve indicates the fluid surface at rest at $t=0$ with a crest at $x=0$ and $\varphi=0$ while the dashed red curve shows the surface at rest at $t=T/2$ with crests at $x=\pm\pi$. The dotted black line shows the flat surface at $t=T/4$ when the fluid celerity is maximal. Thin gray curves show intermediate profiles. Filled dots mark the nodes at $x=\pm\pi/2$. 
The data cover the quarter cell, $0\le x\le\pi$ and $0\le t\le T/4$, and the solution over the full space-time period is reconstructed using \eqref{eq:periodic} and \eqref{eq:sym}.}
  \label{fig:sketch}
\end{figure}

The standing waves considered here are space-time periodic solutions of \eqref{eq:kin}-\eqref{eq:dyn} with an instant at which the fluid is at rest. Writing $T$ for the temporal period, 
\begin{equation}
  u(x+2\pi,t)=u(x,t+T)=u(x,t), \qquad u\in\{\eta,\varphi_x,\varphi_y\}.
  \label{eq:periodic}
\end{equation}
We place an instant of rest at $t=0$ with a crest at $x=0$. We assume without loss of generality $\varphi(x,0)=0$ for all $x$. Additionally, we assume  
\begin{equation}
    \eta(-x,t)=\eta(x,t),\quad \varphi_x(x,t)=-\varphi_x(-x,t)\quad\text{and}\quad \varphi_y(x,t)=\varphi_y(-x,t).
  \label{eq:sym}
\end{equation}
These periodicity and symmetry conditions, together with time reversibility, imply that the solution on the quarter cell, $0\le x\le\pi$ and $0\le t\le T/4$, determines the solution over the full space-time period. Even reflection in $x$ recovers $-\pi\le x\le\pi$, and reflections in $t$ (with sign reversals of $\varphi$ and $\mu$) recover $0\leq t\leq T$. See Figure \ref{fig:sketch} for a schematic. The numerical solutions in  \cite{wilkening2011breakdown,wilkening2012overdetermined} place $t=0$ a quarter period before the rest state. The two conventions are related by $t\mapsto t-T/4$. 

The dataset consists of $140$ solutions, labeled $\mathrm{idx}=101,\ldots,240$, which discretely sample a one-parameter family from small to large amplitudes. Each solution stores $201$ equally spaced snapshots on $[0,T/4]$, with $\eta$, $\varphi$, and $\mu$ sampled at $257$ equally spaced points on $[0,\pi]$. Even reflection, without duplicating the endpoints, gives $512$ uniform points on $(-\pi,\pi]$. 

A companion table records, for each solution in the dataset, the temporal period $T$, and the wave height, defined as
\begin{equation}
  h=\frac12(\eta(0,0)-\eta(\pi,0))
  \label{eq:height}
\end{equation}
(so that $2h$ is the vertical crest-to-trough distance at an instant of rest), together with the acceleration of a fluid particle at the crest at rest, denoted by $A_c$ (see \eqref{eq:Ac}), and the maximal slope measures, $\max_x|\eta_x(\cdot,0)|$ and $\|\eta_x(\cdot,0)\|^2$.

\subsection{Crest acceleration as a parameter}\label{sec:eps}

The wave height in \eqref{eq:height} is a natural candidate for parameterizing standing waves, but it is unfortunately not one-to-one along the dataset. It reaches a local maximum at $\mathrm{idx}=209$ and then decreases slightly, so that many large amplitude members share nearly the same height $h\approx0.616$--$0.618$. The temporal period $T$ also exhibits turning points.

The acceleration of a fluid particle at the crest provides an alternative, physically meaningful parameter. At an instant of rest, the downward acceleration is
\begin{equation*}
  A_c=-\eta_{tt}(0,0).
\end{equation*}
The vertical Euler equation then gives
\begin{equation}
  A_c=g-\rho^{-1}p_y(0,\eta(0,0),0),
  \label{eq:Ac}
\end{equation}
where $p$ is the pressure and $\rho$ is the fluid density. We define
\begin{equation}
  \eps=1-\frac{A_c}{g}
      =\frac{p_y(0,\eta(0,0),0)}{\rho g},
  \label{eq:eps}
\end{equation}
which measures the crest's distance from free fall. Small $\eps$ indicates that the particle acceleration at the crest at rest is nearly gravitational acceleration. Alternatively, $\eps$ measures the upward pressure gradient relative to the gravitational force. Importantly, $\eps$ decreases monotonically across the dataset, from approximately $0.99$ at $\mathrm{idx}=101$ to $0.012$ at $\mathrm{idx}=240$. We do not claim that $\eps$ parametrizes the complete solution set globally.

Over the final segment of the dataset, the crest height changes little, from $0.8217$ to $0.8245$ for the final ten members, whereas the maximal slope has a correlation with $\eps$ greater than $0.9999$ on a $\log$--$\log$ scale. Decreasing $\eps$ therefore tracks an increasingly narrow crest with little change in height. It is noteworthy, however, this observation alone does not guarantee a limiting angle or an asymptotic scaling law. Indeed, the oscillatory behavior found in \cite{wilkening2011breakdown} cautions against a self-similar extrapolation all the way out to $\eps=0$.

At the opposite end as $\eps\to 1$, setting $g=1$ and wavenumber $=1$, the leading-order solution is
\begin{equation}
  \eta(x,t)\approx h\cos x\cos t\quad\text{and}\quad
  \varphi(x,t)\approx-h\cos x\sin t.
  \label{eq:linear}
\end{equation}
Consequently, $A_c\approx h$, $h\approx1-\eps$, and $T\to2\pi$ as $\eps\to1$. 

These two limits of $\eps$ have different characteristics and will play complementary roles in our numerical experiments.

\subsection{Reference computations}\label{sec:standard}

The reference solutions used in the present study were computed by Wilkening \cite{wilkening2011breakdown} using the overdetermined shooting method of \cite{wilkening2012overdetermined}. We parametrize them by $\eps$, so that $\eps\to1$ is the linear limit and $\eps\to0$ the free-fall limit. We summarize the method here because its details and, particularly, its cost inform what a surrogate must accomplish. Throughout, numerical results use nondimensional variables and $g=1$.

We seek to determine the rest surface profile $\eta(\cdot,0)=:\eta_0$ (because $\varphi(\cdot,0)=0$) and the temporal period $T$ by solving \eqref{eq:kin}-\eqref{eq:dyn} as a nonlinear boundary value problem in time, subject to periodicity and symmetry conditions, so that the solution returns to the flat-surface state at $t=T/4$. In the implementation, the unknowns are the Fourier coefficients of $\eta_0$ together with $T$, and a scalar (the crest acceleration $A_c$, in \eqref{eq:Ac}, or a Fourier coefficient of $\eta_0$) is held fixed to select a member of the resulting one-parameter solution family.
A trial pair $(\eta_0,T)$ is advanced from $t=0$ to $T/4$ using a spectral discretization in $x$ and a high-order integration of \eqref{eq:kin}-\eqref{eq:dyn} in $t$, in which $\dno\varphi$ is evaluated at each time step after solving \eqref{eq:mu} for $\mu$. The endpoint conditions,
\begin{equation*}
  \widehat\eta_k(T/4)=0\quad \text{for $k$ odd}\quad\text{and}\quad
  \widehat\varphi_k(T/4)=0\quad \text{for $k$ even},
\end{equation*}
where hats denote spatial Fourier coefficients, are imposed in more components than there are unknowns, giving an overdetermined nonlinear least-squares problem. 
The resulting residual is reduced to round-off accuracy by a Levenberg--Marquardt iteration whose Jacobian is assembled from one integration of the linearized equations per unknown. We emphasize that $T$ is not a prescribed input but an output of the solve. It is four times the time at which the flat-surface state is first reached.

\begin{figure}[ht]
\centering
\includegraphics[width=\textwidth]{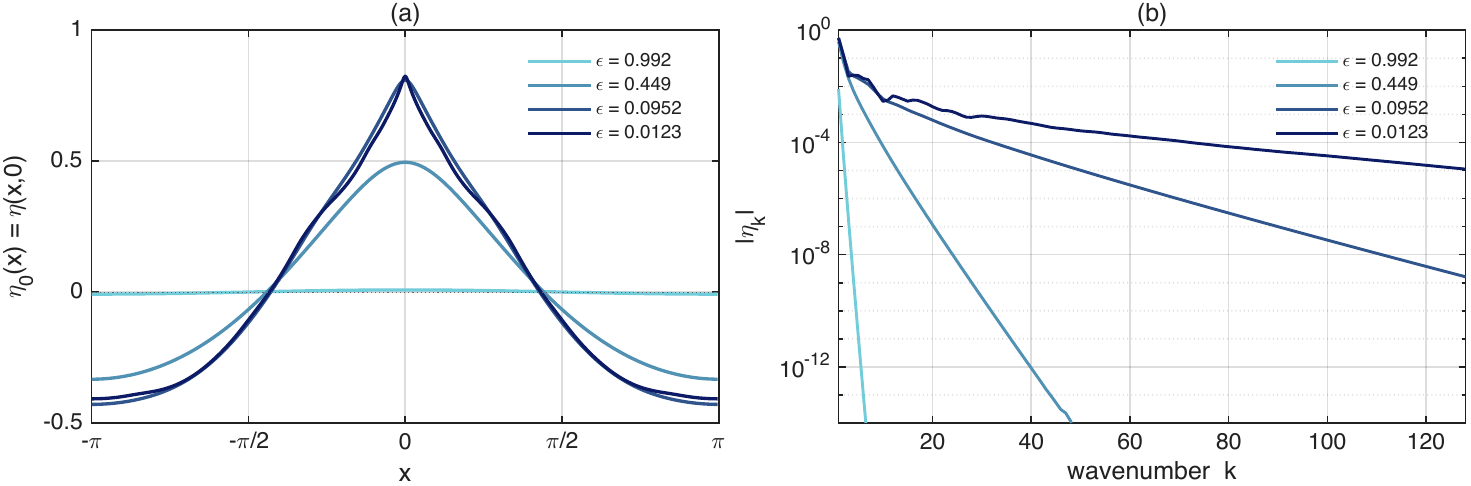}
\caption{
Rest surface profiles and Fourier spectra along the computed solution family. (a)~$\eta_0(x)=\eta(x,0)$ for four members, labeled by $\eps$. The crest sharpens and the trough flattens as $\eps$ decreases. (b)~Fourier coefficient magnitudes $|\widehat{\eta_0}_k|$ of the same solutions. The smallest-amplitude solution is resolved by a few modes, whereas the spectrum of the steepest profile decays slowly and reaches the resolution limit of the $512$-point grid on which the solutions are provided.}
\label{fig:dataset}
\end{figure}

Numerical continuation proceeds along the solution family by stepping the parameter and using the converged pair $(\eta_0,T)$ as the initial guess at the next parameter value. The number of Fourier modes, the number of time steps, and the working precision are increased when needed. Figure \ref{fig:dataset} illustrates why.  

As $\eps$ decreases, the crest of the rest surface profile sharpens and its Fourier spectrum decays more slowly, so that near the free-fall limit, $\eta_0$ carries significant content in many thousands of Fourier modes. Indeed, the small-scale oscillations near the crest can be resolved only in extended-precision arithmetic \cite{wilkening2011breakdown}. 
The cost is dominated by the Jacobian computation. Each Levenberg--Marquardt iteration requires as many quarter-period integrations as there are unknowns, and each integration requires solving \eqref{eq:mu} at each time step. The cost per member increases from seconds near the linear limit to hours near the free-fall limit \cite{wilkening2011breakdown, wilkening2012overdetermined}.

\begin{figure}[h]
  \centering
  \includegraphics[width=\textwidth]{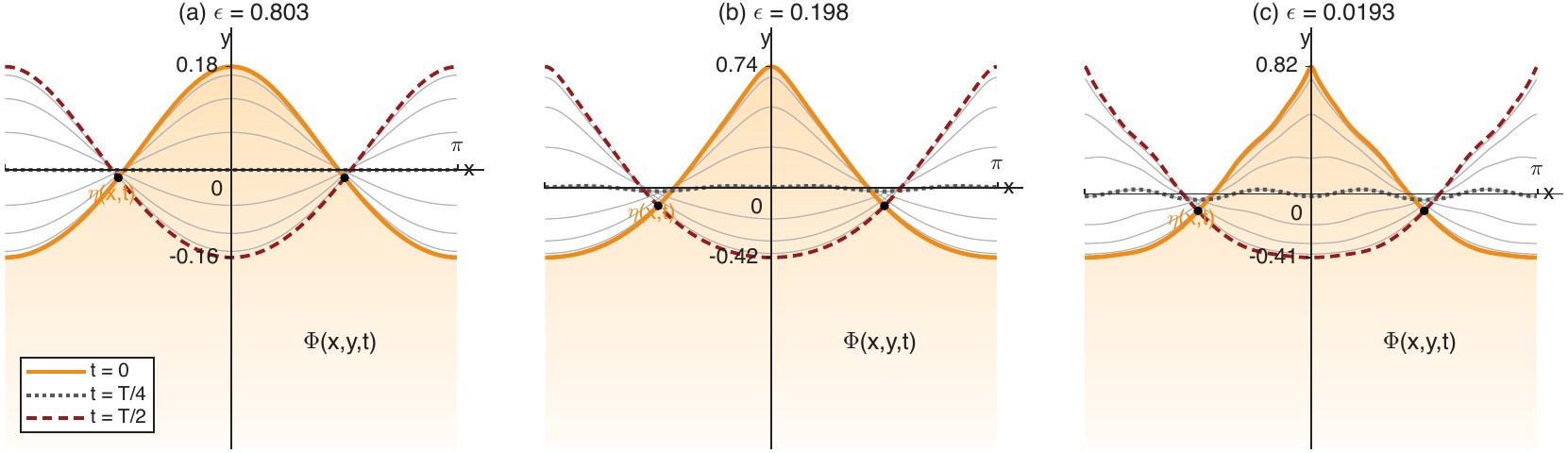}
  \caption{
  Wave profiles for three members of the computed solution family, (a)~$\eps\approx0.80$, (b)~$\eps\approx0.20$, and (c)~$\eps\approx0.019$, shown using the conventions of Figure \ref{fig:sketch}. Solid, dotted, and dashed curves indicate $t=0$, $T/4$, and $T/2$, respectively, and thin gray curves show intermediate profiles. Dots mark the nodes at $x=\pm\pi/2$. (Note that $\eta(\pm\pi/2,t)$ needs not be zero.) The shaded region indicates the fluid at $t=0$. Vertical tick labels give the crest and trough values of $\eta(\cdot,0)$ and differ from panel to panel. The second half of the period retraces the first using $\eta(x,T-t)=\eta(x,t)$.}
  \label{fig:family}
\end{figure}

\begin{figure}[ht]
  \centering
  \includegraphics[width=\textwidth]{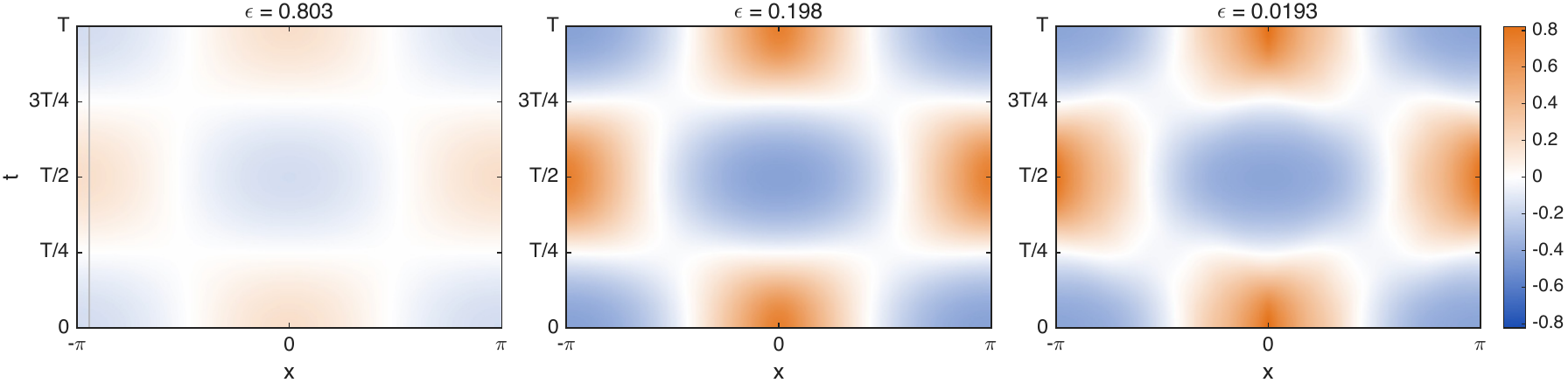}
  \caption{
  $\eta(x,t)$ in space and time over one full period for the three solutions in Figure \ref{fig:family}, using a common color scale. The crest is located at $x=0$ at $t=0,T$ while at $x=\pm\pi$ at $t=T/2$. The surface is flat at $t=T/4$ and $3T/4$. Only the quarter cell, $0\le x\le\pi$ and $0\le t\le T/4$, is computed. The remainder is reconstructed using \eqref{eq:periodic} and \eqref{eq:sym}.}
  \label{fig:spacetime}
\end{figure}

Figures \ref{fig:family} and \ref{fig:spacetime} show three representative members of the computed solution family, from the nearly linear end ($\eps\approx0.8$) to the nearly free-fall end ($\eps\approx0.02$), as sequences of wave profiles using the conventions of Figure \ref{fig:sketch} and as a space-time field over one full period. As $\eps$ decreases along the family, the crest sharpens and the trough flattens. Indeed, as $\eps\to0$, the fluid particle at the crest at rest is momentarily in free fall, which requires the pressure gradient there to vanish and, hence, the surface to be nearly cornered.

Recall that the cost is dominated by the Jacobian computation and increases along the computed solution family. Furthermore, each numerical run yields a solution only at one parameter value, so that the parameter dependence of the solution is known only for the computed members. Additionally, the endpoint of the dataset is where the computation becomes prohibitive \cite{wilkening2011breakdown}. It is not a mathematically established endpoint of the solution set or of the computational reach of conventional methods. These motivate what a surrogate must deliver. 


\section{Learning the solution operator}\label{sec:fno}

\subsection{Solution operator formulation}\label{sec:operator}

In our time convention, $\varphi(\cdot,0)=0$, so that $\eta(\cdot,0)$ determines the subsequent standing wave evolution. The learning target is therefore the solution operator of \eqref{eq:kin}-\eqref{eq:dyn}, restricted to the sampled solution family, which maps a wave profile at an instant of rest and an observation time to the fluid surface, the velocity potential and the dipole density at the surface at that time. For a rest wave profile $\eta_0=\eta(\cdot,0)$ represented in the dataset and its temporal period $T=T(\eta_0)$, we represent the learned operator as
\begin{equation}
  \mathcal{S}_\theta:(\eta_0,\tau)\longmapsto
  (\eta,\varphi,\mu)(\cdot,\tau T/4),
  \qquad \tau\in[0,1],
  \label{eq:operator}
\end{equation}
where $\theta$ is the learned parameters. Here the normalized time $\tau$ measures the fraction of each solution's quarter period, so that all members share the common interval from the rest state ($\tau=0$) to the quarter-period state ($\tau=1$) where the fluid surface is flat. The period is therefore excluded from the learning problem and can be supplied when conversion to physical time is needed. 
We emphasize that \eqref{eq:operator} predicts the full solution state, without projection onto a small modal basis or successive time stepping. Moreover, it evaluates any requested $\tau$ directly, at a cost independent of $\tau$.

A conventional neural network acting on discrete vectors is typically tied to a sampling grid. In neural operator learning, the learned operator acts on functions instead, with the grid entering only through their numerical representation. The Fourier neural operator (FNO) \cite{li2021fourier} implements this in Fourier space and is particularly well-suited to the standing wave problem for several reasons. The spatial domain is periodic and the data lie on a uniform grid. The linear standing wave, $\eta(x,t)=(1-\eps)\cos x\cos t$ and $\varphi(x,t)=(\eps-1)\cos x\sin t$ (see \eqref{eq:linear}), consists of the fundamental harmonic while nonlinear waves contain higher harmonics. Additionally, the Dirichlet-Neumann operator, which encodes the nonlocality of \eqref{eq:kin}-\eqref{eq:dyn}, becomes a Fourier multiplier 
\begin{equation*}
  G(0)\varphi
    =\sum_{k\in\mathbb{Z}}|k|\widehat\varphi_k e^{\mathrm{i}kx}
\end{equation*}
in the flat-surface limit. However, it depends on $\eta$ and no longer has such a constant-multiplier form for a deformed surface.

\subsection{Spectral convolution}\label{sec:specconv}

Convolution becomes multiplication in Fourier space, and a spectral convolution layer learns the Fourier multipliers rather than the kernel. Let $v:\T\to\R^{d}$ denote a $d$-channel function on a periodic grid with discrete Fourier coefficients $\widehat v_k\in\mathbb{C}^{d}$. A spectral convolution layer applies
\begin{equation}
  \widehat{\mathcal{K}v}_k=
  \begin{cases}
    R_k\widehat v_k & \text{for $|k|\le k_{\max}$},\\
    0 & \text{for $|k|>k_{\max}$},
  \end{cases}
  \label{eq:specconv}
\end{equation}
where $R_k\in\mathbb{C}^{d\times d}$ are learned matrices and $k_{\max}$ is a truncation wavenumber. An inverse transform returns the result to physical space. For a real-valued field, the Fourier coefficients have conjugate symmetry, and the learned multipliers are represented with the corresponding symmetry $R_{-k}=\overline{R_k}$.

This spectral convolution layer has two useful properties for the present purposes. First, each retained mode acts globally, coupling points throughout the domain. Particularly, $G(0)$ in the flat-surface limit has the form \eqref{eq:specconv} with $R_k=|k|$. Second, the periodic domain and uniform grid match the data, and no boundary treatment or padding is required.

The spectral path discards wavenumbers above $k_{\max}$. It does not impose a strict band limit on the full network, and higher-frequency content can be generated by the nonlinearity of the layers and pass through the pointwise paths (see \eqref{eq:layer}), just as higher harmonics are generated in the solution by the nonlinearity of \eqref{eq:kin} and \eqref{eq:dyn}. Spectral truncation is expected to have little effect when the Fourier spectrum of the solution decays rapidly, but can limit the representation when the decay is slow. 

\subsection{Network architecture}\label{sec:arch}

\begin{figure}[ht]
\centering
\includegraphics[width=\textwidth]{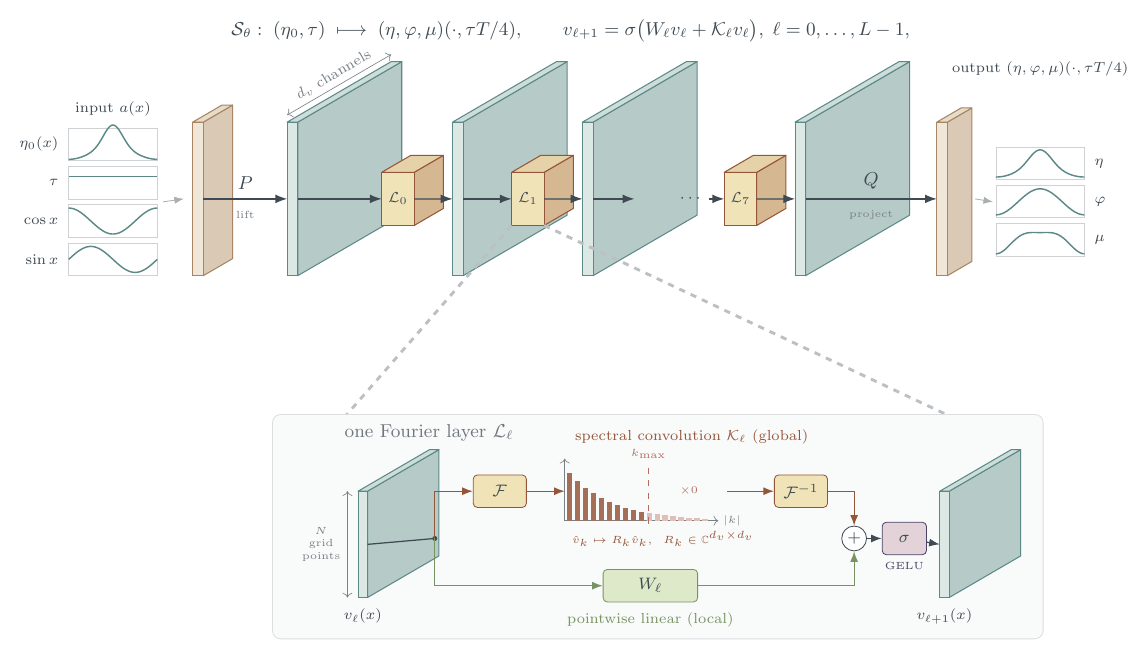}
\caption{Schematic of the learned operator \eqref{eq:operator}. Top: The four-channel input $a(x)=(\eta_0(x),\tau,\cos x,\sin x)$ is lifted by a pointwise map $P$ to $d_v=64$ channels, processed through $L=8$ Fourier layers, and projected by a pointwise map $Q$ to the three output fields $(\eta,\varphi,\mu)(\cdot, \tau T/4)$. 
Bottom: Each Fourier layer combines the spectral convolution $\mathcal{K}_\ell$, in \eqref{eq:specconv}, acting globally on modes $|k|\le k_{\max}=64$, with the pointwise path $W_\ell$, acting locally in physical space, followed by the GELU nonlinearity. See \eqref{eq:layer} for one Fourier layer. 
}
\label{fig:arch}
\end{figure}

The four-channel input of our neural network is
\begin{equation}
  a(x)=(\eta_0(x),\tau,\cos x,\sin x),
  \label{eq:a}
\end{equation}
where the constant channel $\tau$ conditions the operator on the requested time and $(\cos x,\sin x)$ encodes the position on the unit circle continuously across $x=\pm\pi$. These position channels span the fundamental spatial harmonic, and the linear standing wave uses the cosine component.
The network first lifts the input $a$ to a $d_v$-channel function $v_0=P a$ by a pointwise linear map $P$, then applies $L$ Fourier layers
\begin{equation}
  v_{\ell+1}(x)=\sigma(W_\ell v_\ell(x)+(\mathcal{K}_\ell
  v_\ell)(x)),
  \qquad \ell=0,\dots,L-1,
  \label{eq:layer}
\end{equation}
and finally projects $v_L$ to the three-channel output $(\eta,\varphi,\mu)$ by a pointwise map $Q$.
Here, $\mathcal{K}_\ell$ is a spectral convolution layer in \eqref{eq:specconv} with its own multipliers, and $W_\ell$ is a pointwise linear map that combines channels locally in physical space without passing through Fourier space; and $\sigma$ is the Gaussian error linear unit (GELU) nonlinearity.  
The separation between linear nonlocal operations in Fourier space and nonlinear pointwise operations in physical space resembles the pseudo-spectral discretization, although the weights are learned and the layers are not time steps.   
In the numerical experiments, we use $L=8$ layers, a width of $d_v=64$, and a truncation wavenumber of $k_{\max}=64$, out of the $257$ nonnegative wavenumbers representable on the $512$-point grid. Figure \ref{fig:arch} summarizes the architecture. The cutoff is chosen to accommodate the Fourier spectra of rest profiles, such as those in Figure \ref{fig:dataset}(b), and its possible influence on extrapolation is discussed in Section \ref{sec:exp1}.

\section{Numerical experiments}\label{sec:results}

We assess \eqref{eq:operator} through two experiments, extrapolation toward the free-fall limit as $\eps\to0$ (Experiment~I) and toward the linear limit as $\eps\to1$ (Experiment~II). In these experiments, the solution operators are trained independently and evaluated on disjoint in-distribution (ID) and out-of-distribution (OOD) solution sets. The ID solutions lie inside the training range and measure interpolation accuracy while the OOD solutions occupy one end of the reference solution branch and measure extrapolation accuracy. Section \ref{sec:design} describes the data and training procedure, and Section \ref{sec:criteria} defines the prediction errors and a residual of the governing equations. The latter can be evaluated without reference solutions. Sections \ref{sec:exp1} and \ref{sec:exp2} report the prediction errors of the two experiments, and Section \ref{sec:residpred} compares the residuals.

\subsection{Data partitions and training}\label{sec:training}
\label{sec:design}

The data consist of solutions computed at a discrete set $\mathcal{E}\subset(0,1)$ of values of the parameter $\eps$. 
All fields are sampled on the uniform spatial grid
\begin{equation}
  x_i=-\pi+\frac{2\pi i}{N},\qquad i=1,\dots,N,\qquad N=512,
  \label{eq:xgrid}
\end{equation}
and at the normalized time instances
\begin{equation}
  \tau_j=\frac{j-1}{M-1},\qquad j=1,\dots,M,\qquad M=101,
  \label{eq:tgrid}
\end{equation}
which correspond to the physical times $t_j^{\eps}=\tau_jT^\eps/4$ where $T^\eps$ is the temporal period of the solution at $\eps$. These $M=101$ samples correspond to every other one of the $201$ stored time instances described in Section \ref{sec:problem}. The target output of \eqref{eq:operator} at $x=x_i$ and $\tau=\tau_j$ is the three-component state, denoted by
\begin{equation}
  \bm{u}^{\eps}_{j}(x_i)=(\eta,\varphi,\mu)^{\eps}(x_i,\,t_j^{\eps}),
  \label{eq:u_ij}
\end{equation}
and for a single field $u\in \{\eta,\varphi,\mu\}$, we write $u^{\eps}_{ij}=u^{\eps}(x_i,t^{\eps}_j)$ for the corresponding component of $\bm u^{\eps}_j(x_i)$. This notation is used throughout the section. The corresponding input is 
\begin{equation}
    a^{\eps}_j(x_i)=(\eta_0^{\eps}(x_i),\tau_j,\cos x_i,\sin x_i),
    \label{eq:a_ij}
\end{equation}
where $\eta_0^{\eps}$ is the rest surface profile of the solution at $\eps$.

\paragraph{Partition of the data.}
In our experiments, the parameter set is partitioned into three disjoint subsets
\begin{equation}
\mathcal{E}=\mathcal{E}_{\mathrm{tr}}\cup\mathcal{E}_{\mathrm{ID}}
  \cup\mathcal{E}_{\mathrm{OOD}},
  \qquad
  N_{\mathrm{tr}}:=|\mathcal{E}_{\mathrm{tr}}|=125,\quad
  N_{\mathrm{ID}}:=|\mathcal{E}_{\mathrm{ID}}|=5,\quad
  N_{\mathrm{OOD}}:=|\mathcal{E}_{\mathrm{OOD}}|=10,
  \label{eq:partition}
\end{equation}
where the subscripts denote the training set (tr), the in-distribution test set (ID), and the out-of-distribution test set (OOD), and $|\cdot|$ is the number of elements of a set. The values in $\mathcal{E}_{\mathrm{tr}}$ are used to fit the operator while those in $\mathcal{E}_{\mathrm{ID}}$ and $\mathcal{E}_{\mathrm{OOD}}$ are withheld from training and used only for evaluation. The training set is
\begin{equation*}
  \mathcal{D}_{\mathrm{tr}}=\{(a^{\eps}_j,\bm{u}^{\eps}_j):
  \eps\in\mathcal{E}_{\mathrm{tr}},j=1,\dots,M\},
\end{equation*}
where $a^{\eps}_j$ and $\bm{u}^{\eps}_j$ are in \eqref{eq:a_ij} and \eqref{eq:u_ij}, respectively, and are sampled at the $N$ grid points in \eqref{eq:xgrid}. The test sets $\mathcal{D}_{\mathrm{ID}}$ and $\mathcal{D}_{\mathrm{OOD}}$ are defined in the same manner with $\mathcal{E}_{\mathrm{ID}}$ and $\mathcal{E}_{\mathrm{OOD}}$ in place of $\mathcal{E}_{\mathrm{tr}}$. 

For each $\eps\in\mathcal{E}_{\mathrm{ID}}$, the nearest smaller and the nearest larger values of $\eps$ in $\mathcal{E}$ belong to $\mathcal{E}_{\mathrm{tr}}$, so that no two ID values of $\eps$ are adjacent and no ID value is adjacent to the OOD block. The ID error therefore measures the accuracy with which the operator interpolates between neighboring training solutions. The set $\mathcal{E}_{\mathrm{OOD}}$ consists of either the $N_{\mathrm{OOD}}$ smallest or the $N_{\mathrm{OOD}}$ largest values of $\eps$ in $\mathcal{E}$, depending on the experiment, so that each OOD value lies outside the interval $[\min\mathcal{E}_{\mathrm{tr}},\,\max\mathcal{E}_{\mathrm{tr}}]$ spanned by the training values. The OOD error therefore measures the accuracy of extrapolation toward the corresponding limit $\eps\to0$ or $\to1$.

\paragraph{Experiment I: toward the free-fall limit.}

The parameter sets satisfy
\begin{equation*}
  \mathcal{E}_{\mathrm{tr}}\subset[0.030,\,0.992],\qquad
  \mathcal{E}_{\mathrm{ID}}\subset[0.061,\,0.87],\qquad  \mathcal{E}_{\mathrm{OOD}}\subset[0.012,\,0.028],
\end{equation*}
so that $\max\mathcal{E}_{\mathrm{OOD}}<\min\mathcal{E}_{\mathrm{tr}}$. Indeed, the OOD set consists of the $N_{\mathrm{OOD}}$ smallest values in $\mathcal{E}$.
Defining the partition in terms of $\eps$ is essential. Recall that the wave height, in \eqref{eq:height}, is not monotone along the reference solution branch. Particularly, the OOD solutions share values of $h$ with training solutions on the opposite side of the maximum of $h$, so that they would not lie outside the training range in terms of $h$. Their $\eps$ values, on the other hand, lie strictly below $\min\mathcal{E}_{\mathrm{tr}}$.

\paragraph{Experiment II: toward the linear limit.}

The roles of the two ends of the branch are exchanged, and
\begin{equation*}
  \mathcal{E}_{\mathrm{tr}}\subset[0.012,\,0.904],\qquad
  \mathcal{E}_{\mathrm{ID}}\subset[0.039,\,0.76],\qquad
  \mathcal{E}_{\mathrm{OOD}}\subset[0.913,\,0.992],
\end{equation*}
so that $\max\mathcal{E}_{\mathrm{tr}}<\min\mathcal{E}_{\mathrm{OOD}}$. Indeed, the OOD set consists of the $N_{\mathrm{OOD}}$ largest values in $\mathcal{E}$. 

\paragraph{Normalization and training.}

Each output channel is standardized by the mean and standard deviation of the channel over the training set,
\begin{equation}
  m_c=\frac{1}{N_{\mathrm{tr}}MN}\sum_{\eps\in\mathcal{E}_{\mathrm{tr}}}
  \sum_{j=1}^{M}\sum_{i=1}^{N}u^{\eps}_{c,j}(x_i)
  \quad\text{and}\quad
  s_c^2=\frac{1}{N_{\mathrm{tr}}MN}\sum_{\eps\in\mathcal{E}_{\mathrm{tr}}}
  \sum_{j=1}^{M}\sum_{i=1}^{N}\bigl(u^{\eps}_{c,j}(x_i)-m_c\bigr)^2,
  \label{eq:stdz}
\end{equation}
where $c\in \{\eta,\varphi,\mu\}$. The same constants $m_c$ and $s_c$ are applied to the ID and OOD sets. The parameters $\theta$ of the operator are obtained by minimizing the mean squared error of the standardized fields,
\begin{equation}
  \mathcal{L}(\theta)=\frac{1}{3N_{\mathrm{tr}}MN}
\sum_{\eps\in\mathcal{E}_{\mathrm{tr}}}\sum_{j=1}^{M}\sum_{i=1}^{N}
  \sum_{c\in \{\eta,\varphi,\mu\}}\left(\frac{[\fno(a^{\eps}_j)]_c(x_i)-u^{\eps}_{c,j}(x_i)}{s_c}\right)^{2},
  \label{eq:loss}
\end{equation}
with the Adam optimizer over $1000$ epochs, using an initial learning rate of {$10^{-3}$} that decays according to a cosine schedule.
In each experiment, the network parameters are initialized independently, using a fixed random seed that also determines the ordering of the training data. Training and evaluation are both performed on the grid \eqref{eq:xgrid} and \eqref{eq:tgrid}.

\paragraph{Roles of the test sets.}

\begin{figure}[ht]
\centering
\includegraphics[width=\textwidth]{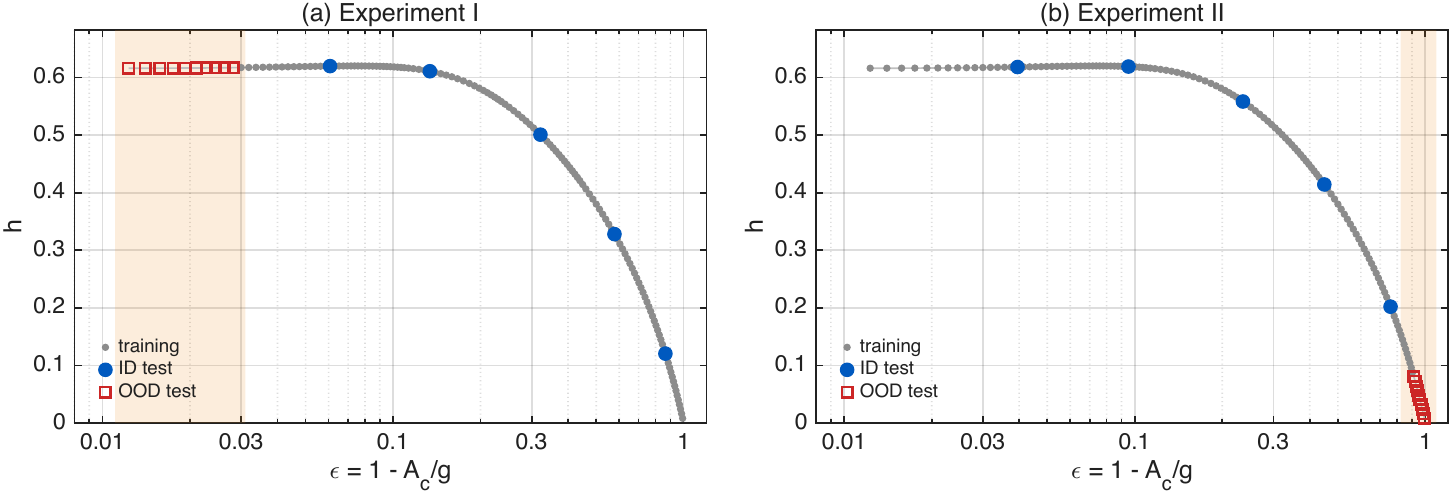}
\caption{
Data partitions \eqref{eq:partition} for (a)~Experiment~I and (b)~Experiment~II: $N_{\mathrm{tr}}=125$ training solutions, $N_{\mathrm{ID}}=5$ ID test solutions, and $N_{\mathrm{OOD}}=10$ OOD test solutions. The wave height $h$ is plotted against $\eps$ on a logarithmic $\eps$ axis. Gray dots denote training solutions, filled circles denote ID solutions, and open squares and shading identify the OOD block. The turning point of $h$ appears at the left of both panels. In Experiment~I, the $h$ values of the OOD solutions overlap those of training whereas their $\eps$ values lie strictly outside the training range.}
\label{fig:splits}
\end{figure}

The ID error measures the accuracy of the operator for values of $\eps$ within the range of the training parameter values, and the comparison of the OOD error with the ID error quantifies the loss of accuracy when extrapolating toward a limiting regime. The ID sets of the two experiments have no values in common and alternate along the branch. Both are shown in Figures \ref{fig:splits} and \ref{fig:fnoresid}.

\subsection{Error measures and a residual}\label{sec:criteria}

Two complementary evaluation measures are used throughout. The first is the relative $L^2$ error of a prediction against the reference solution. The second is a residual of the governing equations. The residual requires no reference solution and can therefore be applied both to the ID and OOD predictions, for which reference solutions exist, and to the exploratory predictions of Section \ref{sec:epstry}, for which they do not.

\paragraph{Relative $L^2$ error.}

For $\eps\in\mathcal{E}$ and $u\in\{\eta,\varphi,\mu\}$, let $u^{\eps}_{ij}=u^{\eps}(x_i,t^{\eps}_j)$, $1\leq i\leq N$ and $1\leq j\leq M$, denote the reference values on the grid \eqref{eq:xgrid}-\eqref{eq:tgrid}, and $\widehat u^{\eps}_{ij}$ the corresponding values predicted by \eqref{eq:operator}. The relative $L^2$ error over the quarter period is
\begin{equation}
  E^{\eps}(u)=
  \left(\frac{\sum_{j=1}^{M}\sum_{i=1}^{N}
        |\widehat u^{\eps}_{ij}-u^{\eps}_{ij}|^{2}}
       {\sum_{j=1}^{M}\sum_{i=1}^{N}|u^{\eps}_{ij}|^{2}}\right)^{1/2}
  \approx
  \frac{\|\widehat u^{\eps}-u^{\eps}\|_{L^2(\T\times(0,T_\eps/4))}}
       {\|u^{\eps}\|_{L^2(\T\times(0,T_\eps/4))}},
  \label{eq:relL2}
\end{equation}
where the quadrature weights of the uniform grid cancel in the ratio. For a set of test values $\mathcal{E}_{te}\in\{\mathcal{E}_{\mathrm{ID}}, \mathcal{E}_{\mathrm{OOD}}\}$, the average error is obtained by extending the sums over $\mathcal{E}_{te}$ before the ratio is taken,
\begin{equation}
  E_{te}(u)=
  \left(\frac{\sum_{\eps\in\mathcal{E}_{te}}\sum_{j=1}^{M}\sum_{i=1}^{N}
        |\widehat u^{\eps}_{ij}-u^{\eps}_{ij}|^{2}}
       {\sum_{\eps\in\mathcal{E}_{te}}\sum_{j=1}^{M}\sum_{i=1}^{N}
        |u^{\eps}_{ij}|^{2}}\right)^{1/2},
  \label{eq:relL2group}
\end{equation}
so that each solution contributes in proportion to its squared norm and the largest amplitude solutions dominate the average measure. The error over all fields is defined as in \eqref{eq:relL2group} with the sums extended, in addition, over $u\in\{\eta,\varphi,\mu\}$. Where the amplitude of the reference solution varies strongly across $\mathcal{E}_{te}$, we also report the absolute root-mean-square error
\begin{equation}
  A^{\eps}(u)=\left(\frac{1}{MN}\sum_{j=1}^{M}\sum_{i=1}^{N}
  \bigl|\hat u^{\eps}_{ij}-u^{\eps}_{ij}\bigr|^{2}\right)^{1/2}
  \label{eq:absrms}
\end{equation}
in the nondimensional units of Section \ref{sec:problem}, to distinguish changes in prediction accuracy from changes in the reference amplitude.

\paragraph{Residual of the governing equations.} \label{sec:residdef}

Both the reference solutions and predictions provide $\eta$ and $\varphi$ on the space-time grid. Evaluating $\dno\varphi$ independently would require solving Laplace's equation. For an exact solution, however, \eqref{eq:kin} gives $\dno\varphi=\eta_t$. Substituting this into \eqref{eq:dyn} yields a residual of \eqref{eq:kin}-\eqref{eq:dyn} that depends only on the two surface fields
\begin{equation}
  R = \varphi_t + g\eta + \frac12\varphi_x^{2} - \frac{(\eta_t+\eta_x\varphi_x)^{2}}
             {2(1+\eta_x^{2})}.
  \label{eq:resid}
\end{equation}
This tests a necessary consequence of the governing equations. 
The residual does not involve $\mu$, which is assessed only through \eqref{eq:relL2}.

The derivatives in \eqref{eq:resid} are discretized on the grid of the data. For the solution at $\eps$ and $u\in\{\eta,\varphi\}$, the $x$-derivative at each time $t^{\eps}_j$ is computed spectrally as
\begin{equation}
  (u^\eps_x)_{\,\cdot\,j} = \mathcal{F}^{-1}\!(\mathrm{i}k
  \mathcal{F}u^\eps_{\,\cdot\,j}),
  \qquad k=-\frac{N}{2}+1,\dots,\frac{N}{2},
  \label{eq:discx}
\end{equation}
where $\mathcal{F}$ is the discrete Fourier transform on the grid \eqref{eq:xgrid} and the coefficient of the mode $k=N/2$ is set to zero. The $t$-derivative is computed by the fourth-order centered difference as
\begin{equation}
  (u^\eps_t)_{ij} = \frac{u^\eps_{i,j-2}-8 u^\eps_{i,j-1}+8 u^\eps_{i,j+1}-u^\eps_{i,j+2}}
                        {12 \Delta t^\eps},
  \qquad 3\le j\le M-2,
  \qquad \Delta t^\eps=\frac{T^\eps}{4(M-1)},
  \label{eq:disct}
\end{equation}
together with the second-order formulas,
\begin{equation}
  (u^\eps_t)_{i1}=\frac{-3u^\eps_{i1}+4u^\eps_{i2}-u^\eps_{i3}}{2\Delta t^\eps}
  \quad\text{and}\quad
  (u^\eps_t)_{i2}=\frac{u^\eps_{i3}-u^\eps_{i1}}{2\Delta t^\eps},
  \label{eq:disctb}
\end{equation}
at the first two time levels, with the corresponding reflected formulas at $j=M-1$ and $M$. Writing $R^{\eps}_{ij}$ for the resulting values of \eqref{eq:resid}, the relative residual of the solution at $\eps$ is
\begin{equation}
  r^{\eps} =
\left(\frac{\sum_{j=1}^{M}\sum_{i=1}^{N}(R^{\eps}_{ij})^{2}}
       {\sum_{j=1}^{M}\sum_{i=1}^{N}
        (((\varphi^\eps_t)_{ij})^{2} +g^{2}\bigl(\eta^{\eps}_{ij})^{2})}\right)^{1/2},
  \label{eq:relresid}
\end{equation}
that is, the root-mean-square value of $R$ divided by the root-mean-square magnitude of the two linear terms $\varphi_t$ and $g\eta$, which dominate the balance \eqref{eq:resid}. The average residual $r_{te}$ over a set $\mathcal{E}_{te}\in \{\mathcal{E}_{\rm ID}, \mathcal{E}_{\rm OOD}\}$ is defined as in \eqref{eq:relL2group} by extending both sums over $\mathcal{E}_{te}$ before the ratio is taken.

\paragraph{Discretization baseline for the residual.}\label{sec:residres}

\begin{figure}[h]
\centering
\includegraphics[width=0.62\textwidth]{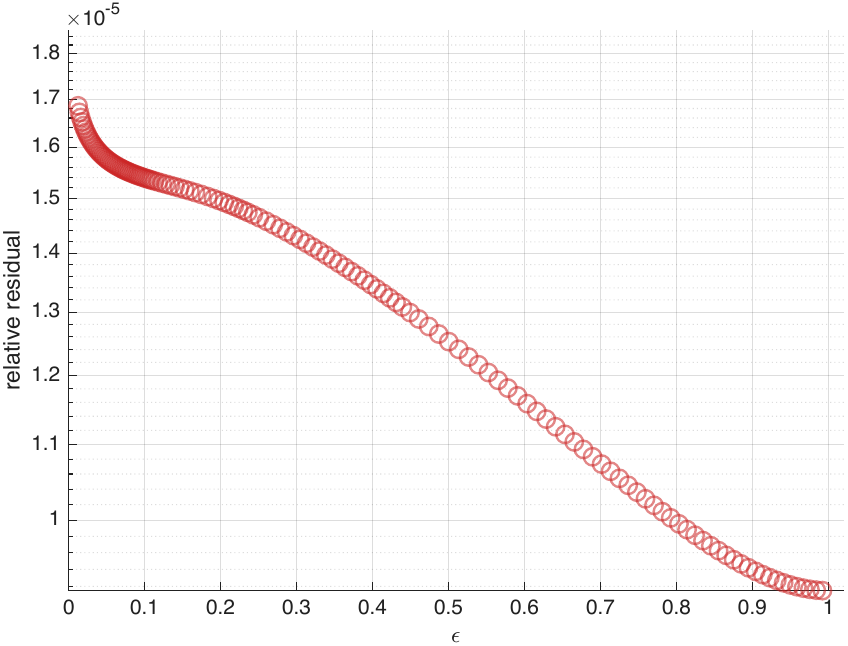}
\caption{
Relative residual \eqref{eq:relresid} of the reference solutions versus $\eps$, evaluated at the same $M=101$ time levels used for the predictions. Values range from $0.9\times10^{-5}$ to $1.7\times10^{-5}$, with a mean of $1.35\times10^{-5}$. The variation along the branch reflects the truncation error of the time difference, rather than a loss of accuracy in the reference data.}
\label{fig:dataresid}
\end{figure}

Even an exact solution generally gives a nonzero residual because of a truncation error in the time differences \eqref{eq:disct} and \eqref{eq:disctb}. Evaluating the residual at the reference solutions with the same formulas and grid used for the predictions provides a baseline for $r^{\eps}$. For the reference solutions of \cite{wilkening2011breakdown}, $r^{\eps}$ lies between $0.9\times10^{-5}$ and $1.7\times10^{-5}$ for every $\eps\in\mathcal{E}$, with a mean of $1.35\times10^{-5}$. See Figure \ref{fig:dataresid}. Accordingly, a prediction with $r^{\eps}$ of order $10^{-5}$ would be indistinguishable from the reference solutions by this diagnostic.

\subsection{Experiment I: toward the free-fall limit}\label{sec:exp1}

Table \ref{tab:err} reports the average relative $L^2$ errors $E_{te}$, in \eqref{eq:relL2group}, for Experiment~I on the ID test set $\mathcal{E}_{te}=\mathcal{E}_{\mathrm{ID}}$ and the OOD test set $\mathcal{E}_{\mathrm{OOD}}$. The three fields have comparable ID errors, ranging from $5.8\times10^{-4}$ to $6.8\times10^{-4}$. This suggests that the operator achieves relative errors below $10^{-3}$ within the training range. Extrapolation to the OOD set increases the error by approximately a factor of five for $\eta$ and three for $\varphi$ and $\mu$. Thus, $\eta$ is the field most affected by extrapolation toward the free fall limit, which is consistent with the sharpening of the crest toward the limit. Both the OOD error of $\eta$ and the combined OOD error exceed the individual errors of $\varphi$ and $\mu$ whereas the three field-wise ID errors remain close.

\begin{table}[ht]
\centering
\caption{Experiment~I. Average relative $L^2$ errors over
the quarter period for the ID test set ($N_{\mathrm{ID}}=5$, $\eps\in[0.061,\,0.87]$) and OOD test set ($N_{\mathrm{OOD}}=10$, $\eps\in[0.012,\,0.028]$).}
\label{tab:err}
\begin{tabular}{lcc}
\toprule
Field & ID & OOD  \\
\midrule
$\eta$     & $6.81\times10^{-4}$ & $3.58\times10^{-3}$ \\
$\varphi$  & $5.79\times10^{-4}$ & $1.95\times10^{-3}$ \\
$\mu$      & $6.36\times10^{-4}$ & $2.02\times10^{-3}$  \\
All fields & $6.35\times10^{-4}$ & $2.35\times10^{-3}$ \\
\bottomrule
\bottomrule
\end{tabular}
\end{table}

\begin{figure}[htbp]
\centering
\begin{subfigure}[b]{.9\textwidth}
  \centering
  \includegraphics[width=\textwidth]{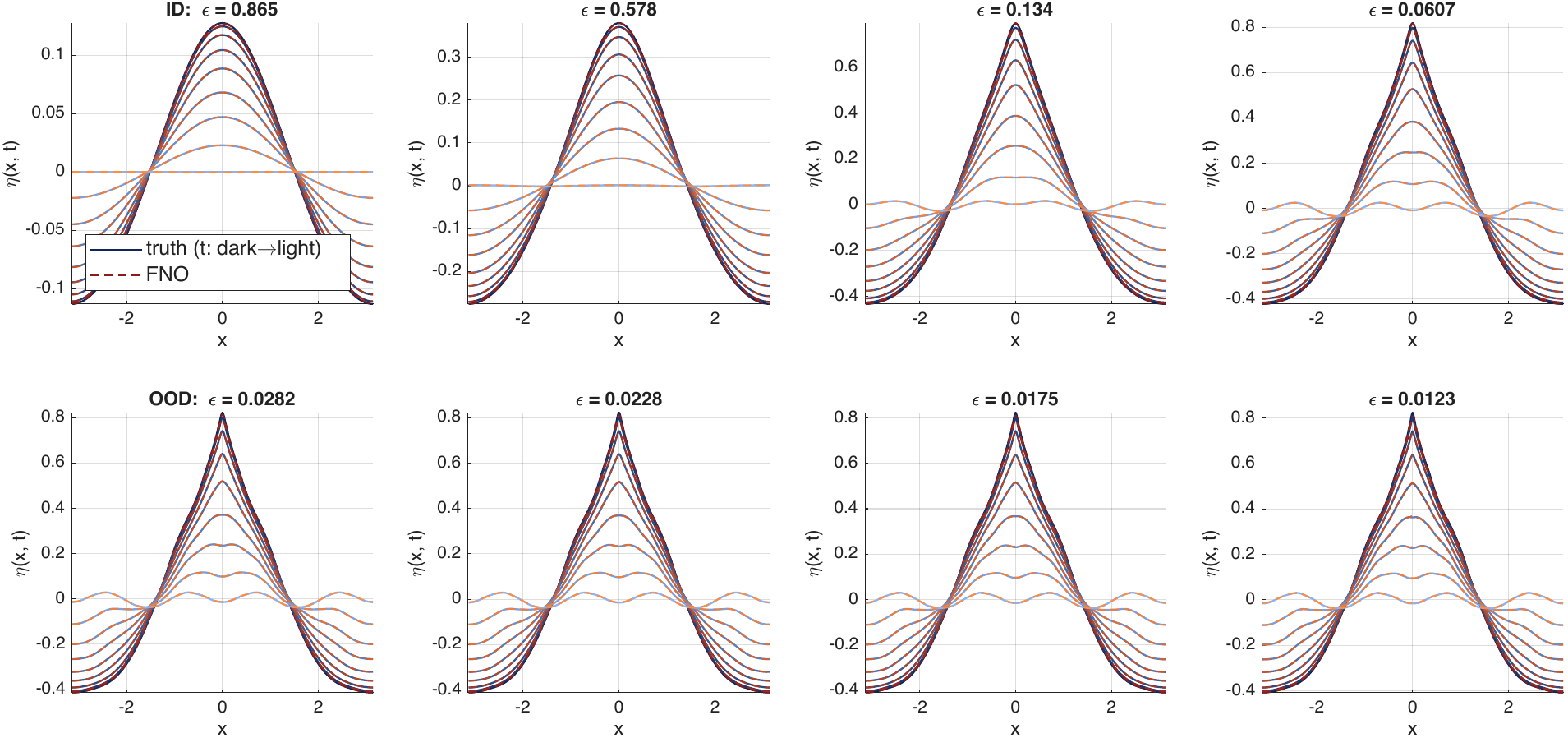}
  \caption{Surface profiles $\eta(x,t)$ over the quarter period for four ID solutions (top row) and four OOD solutions (bottom row). Solid blue curves show the reference solutions, with color progressing from dark to light in time, while dashed red curves show the FNO predictions.}
  \label{fig:exp1-fans}
\end{subfigure}\\
\begin{subfigure}[b]{0.5\textwidth}
  \centering
  \includegraphics[width=\textwidth]{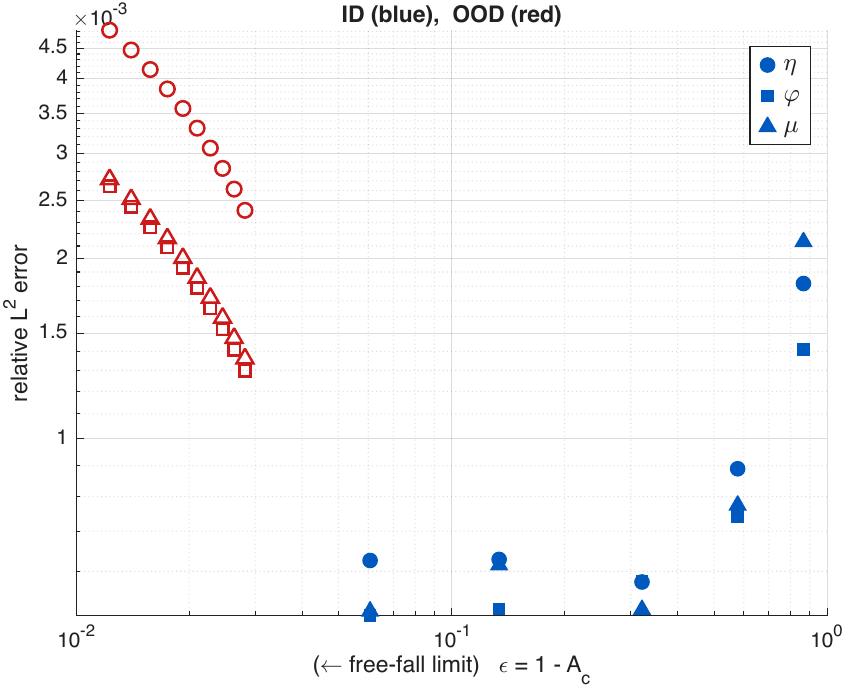}
  \caption{Field-wise relative $L^2$ errors versus $\eps$ for all $\eps\in\mathcal{E}_{\mathrm{ID}}\cup\mathcal{E}_{\mathrm{OOD}}$. Filled and open markers denote ID and OOD solutions, respectively.}
  \label{fig:exp1-error}
\end{subfigure}\\
\begin{subfigure}[b]{.9\textwidth}
  \centering
  \includegraphics[width=\textwidth]{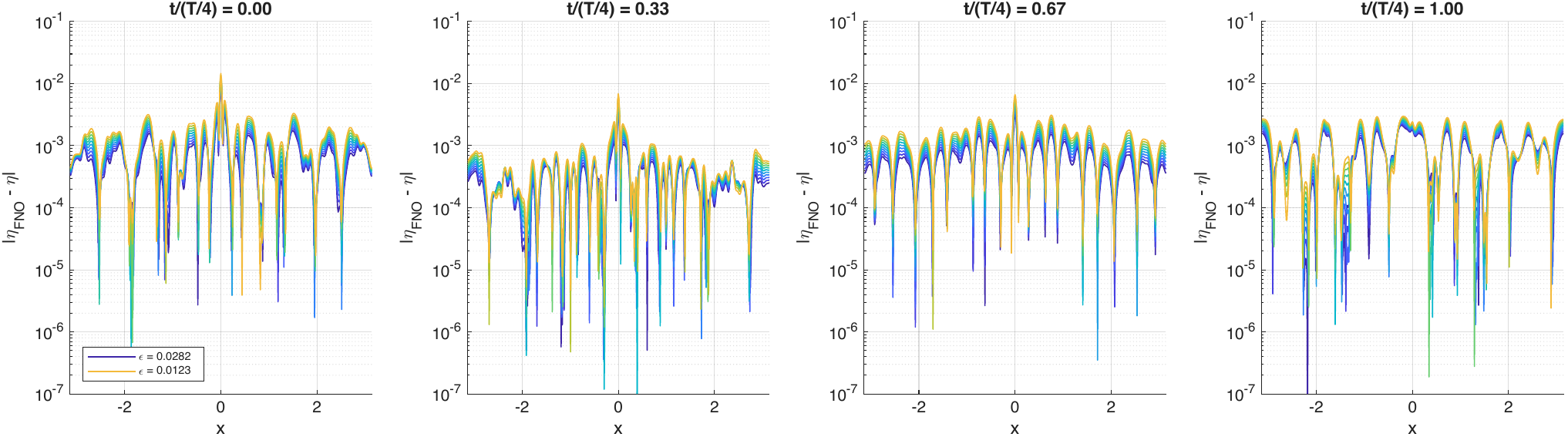}
  \caption{Rest surface profiles near the crest, $|x|\le0.6$, for the solutions in panel (a).}
  \label{fig:exp1-crest}
\end{subfigure}
\caption{Experiment~I. ID and OOD predictions for extrapolation toward the free-fall limit as $\eps\to0$.}
\label{fig:exp1}
\end{figure}

Figure \ref{fig:exp1} resolves these errors for individual solutions. 
Figure \ref{fig:exp1-fans} shows that the sequences of the predicted surface profiles are indistinguishable from the reference solutions at the plotting scale throughout the ID solutions and away from the crest for the OOD solutions. The discrepancies in Table \ref{tab:err} are concentrated near $x=0$ and $t=0$, where the profile is steepest, rather than appearing as a global deformation of the profile. 
Figure \ref{fig:exp1-error} shows that the field-wise errors $E^{\eps}(u)$, $u\in\{\eta,\varphi,\mu\}$, in \eqref{eq:relL2}, has no systematic trend across the ID range $\eps\in[0.061,\,0.87]$, suggesting that interpolation accuracy is governed primarily by the approximation capability of the operator within the training range, rather than by the position along the branch. In contrast, all three OOD errors increase monotonically as $\eps$ decreases, and the gap between the error of $\eta$ and those of $\varphi$ and $\mu$ widens as $\eps\to0$. 
Figure \ref{fig:exp1-crest} shows that the reference crest height remains nearly constant as the crest itself becomes sharper across the OOD set, while the predictions round the crest by the amount that increases with its steepness. The error is therefore localized both in space, near the crest, and in time, near the instant of maximal steepness. 

The dependence of the OOD error on $\eps$ quantifies the loss of accuracy. The error $E^{\eps}(\eta)$ increases monotonically across the OOD set, from $2.41\times10^{-3}$ at $\eps=0.028$ to $4.82\times10^{-3}$ at $\eps=0.012$, and follows closely a power law in $\eps$ over the range. In view of the fine-scale oscillations in the profile along the branch as $\eps\to0$ \cite{wilkening2011breakdown}, we regard this as a local trend over the OOD range rather than an asymptotic law. The reference data suggest a spectral explanation. For the ID solutions, the Fourier coefficients of the rest profile have decayed to negligible values well before the truncation wavenumber $k_{\max}=64$, so that the spectral truncation discards negligible content.
For the steepest OOD solutions, on the other hand, the spectrum decays much more slowly and retains appreciable content above $k_{\max}$, which the spectral convolution, in \eqref{eq:specconv}, cannot represent directly. Across the same OOD set, the crest height $\eta^{\eps}(0,0)$ increases only from $0.8217$ to $0.8245$ while the maximal slope $\max_x|\eta^{\eps}_x(x,0)|$ increases from $0.95$ to $1.13$. As discussed in Section \ref{sec:eps}, the dominant geometric change is crest sharpening in the free-fall limit, and the OOD error is therefore consistent with limited crest resolution. 

\subsection{Experiment II: toward the linear limit}\label{sec:exp2}

Experiment~II tests extrapolation toward the linear limit and reveals a different error pattern. Table \ref{tab:err2} reports the average relative $L^2$ errors $E_{te}$, in \eqref{eq:relL2group}, for Experiment~II on the ID test set $\mathcal{E}_{\mathrm{ID}}$ and the OOD test set $\mathcal{E}_{\mathrm{OOD}}$. The field-wise ID errors, between $5.9\times10^{-4}$ and $7.3\times10^{-4}$, are comparable to those of Experiment~I. On the other hand, the average OOD errors are substantially larger than those of Experiment~I, approximately three times larger for $\eta$ and more than an order of magnitude larger for $\varphi$, $\mu$, and the combined fields. Since each solution contributes to $E_{te}$ in proportion to its squared norm, the largest amplitude OOD solutions at the smallest values of the $\eps$ range dominate these averages. The average measure consequently understates the relative error closer to the linear limit. The individual error $E^{\eps}(\eta)$, in \eqref{eq:relL2}, increases from $4.5\times10^{-3}$ at $\eps=0.913$ to $1.16\times10^{-1}$ at $\eps=0.992$.

\begin{table}[ht]
\centering
\caption{Experiment~II. Average relative $L^2$ errors over
the quarter period for the ID test set ($N_{\mathrm{ID}}=5$ and $\eps\in[0.039,0.76]$) and OOD test set ($N_{\mathrm{OOD}}=10$ and $\eps\in[0.913,0.992]$).}
\label{tab:err2}
\begin{tabular}{lcc}
\toprule
Field & ID & OOD \\
\midrule
$\eta$     & $7.3\times10^{-4}$ & $1.23\times10^{-2}$ \\
$\varphi$  & $5.9\times10^{-4}$ & $3.94\times10^{-2}$ \\
$\mu$      & $6.1\times10^{-4}$ & $4.10\times10^{-2}$ \\
All fields & $6.3\times10^{-4}$ & $3.75\times10^{-2}$  \\
\bottomrule
\end{tabular}
\end{table}

\begin{figure}[htbp]
\centering
\begin{subfigure}[b]{0.98\textwidth}
  \centering
  \includegraphics[width=\textwidth]{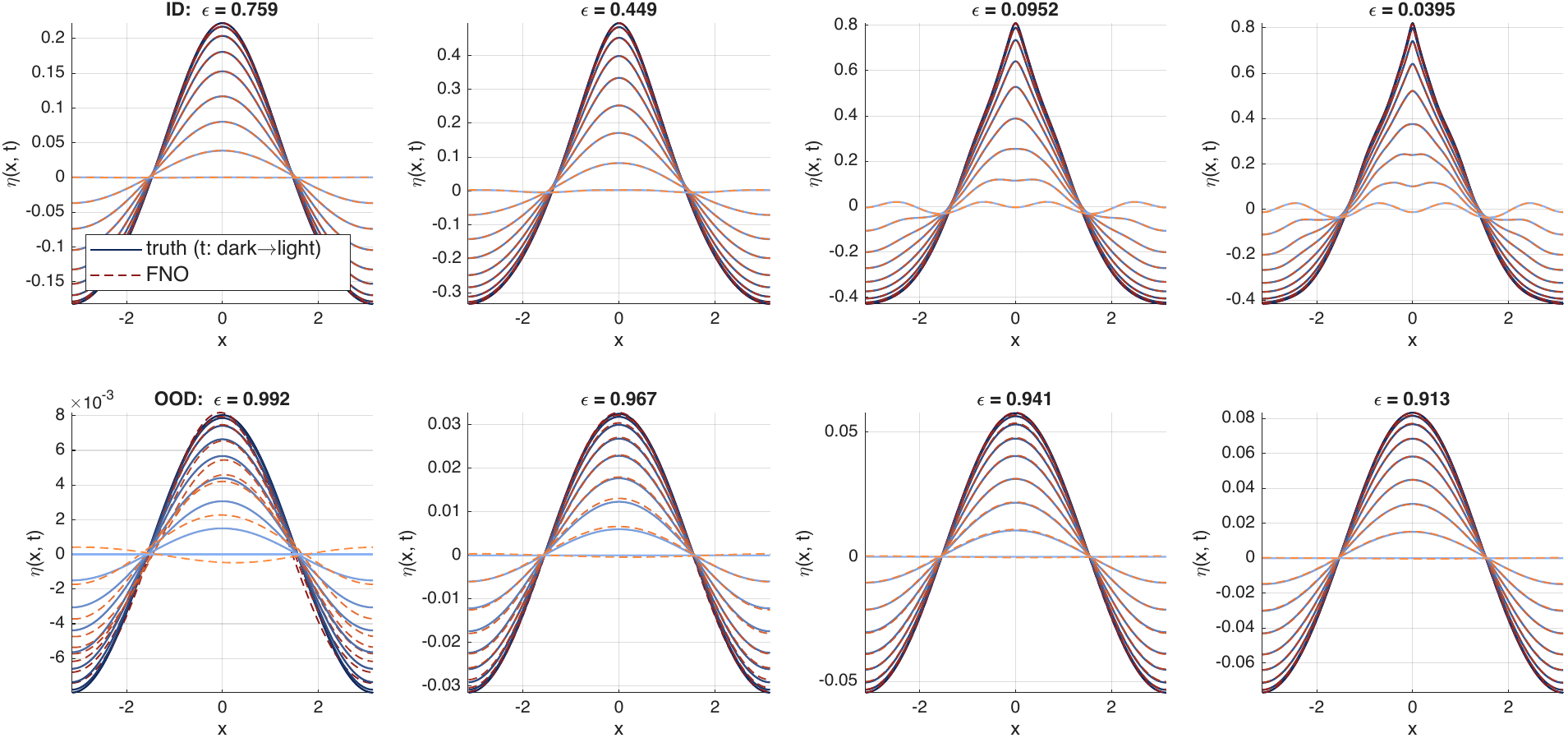}
  \caption{Surface profiles $\eta(x,t)$ over the quarter period for four ID solutions (top row) and four OOD solutions (bottom row). Solid blue curves show the reference solutions, with color progressing from dark to light in time, and dashed red curves show the FNO predictions.}
  \label{fig:exp2-fans}
\end{subfigure}\\
\begin{subfigure}[b]{0.5\textwidth}
  \centering
  \includegraphics[width=\textwidth]{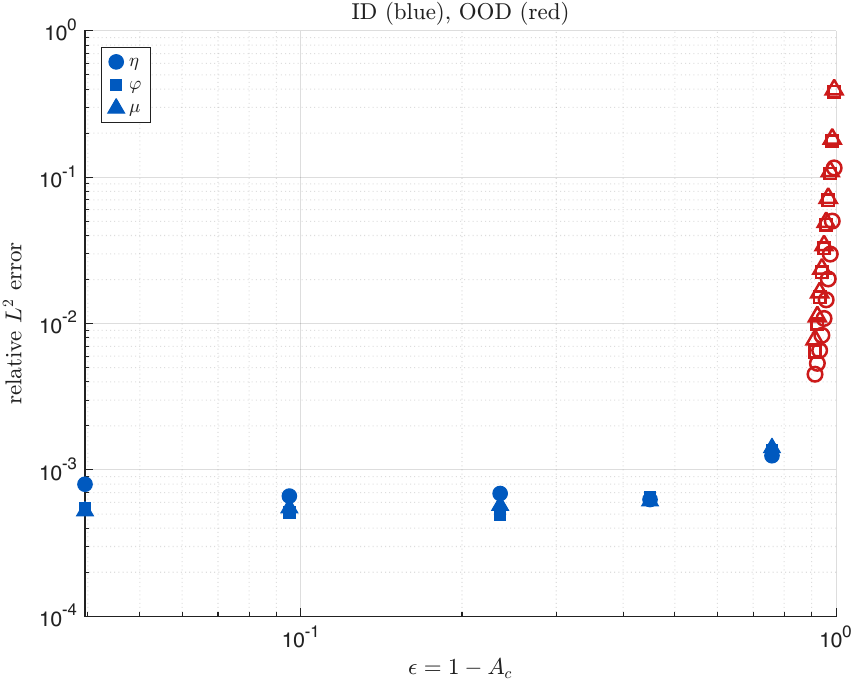}
  \caption{Field-wise relative $L^2$ errors and absolute RMS errors versus $\eps$ for all $\eps\in\mathcal{E}_{\mathrm{ID}}\cup\mathcal{E}_{\mathrm{OOD}}$. Filled and open markers denote ID and OOD solutions, respectively.}
  \label{fig:exp2-error}
\end{subfigure}\\
\begin{subfigure}[b]{.9\textwidth}
  \centering
  \includegraphics[width=\textwidth]{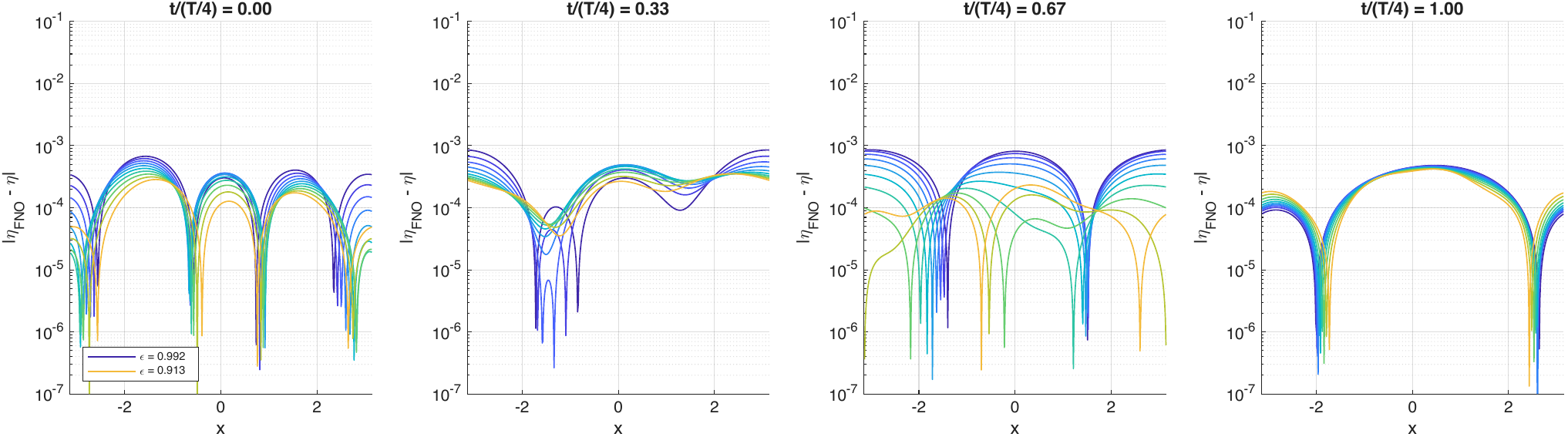}
  \caption{Rest surface profiles near the crest, $|x|\le0.6$, for the solutions in panel (a).}
  \label{fig:exp2-crest}
\end{subfigure}
\caption{Experiment~II. ID and OOD predictions for extrapolation toward the linear limit as $\eps\to1$.}
\label{fig:exp2}
\end{figure}

Figure \ref{fig:exp2} illustrates why the relative errors increase. The surface profile sequences in Figure \ref{fig:exp2-fans} and the crest profiles in Figure \ref{fig:exp2-crest} show that the OOD solutions are nearly sinusoidal with amplitudes that decrease toward the linear limit. The predictions track these profiles closely in absolute terms. Indeed, Figure \ref{fig:exp2-error} shows that the absolute root-mean-square error $A^{\eps}(\eta)$, in \eqref{eq:absrms}, decreases across the OOD set, from $4.6\times10^{-4}$ to $1.8\times10^{-4}$, and remains at or below the ID absolute error, while the relative error $E^{\eps}(\eta)$ increases sharply. The absolute accuracy of the operator therefore does not deteriorate toward the linear limit. On the other hand, the wave amplitude decreases in proportion to $1-\eps$ (see Section \ref{sec:eps}), faster than the absolute error, so that their ratio increases as $\eps\to1$. This is consistent with the standardization in \eqref{eq:stdz}. The larger amplitude training solutions determine the normalization constants, and the loss \eqref{eq:loss} gives no additional weight to the small absolute discrepancies that become significant for small amplitudes.

\subsection{Residual analysis of ID and OOD predictions}
\label{sec:residpred}

\begin{figure}[ht]
\centering
\includegraphics[width=0.62\textwidth]{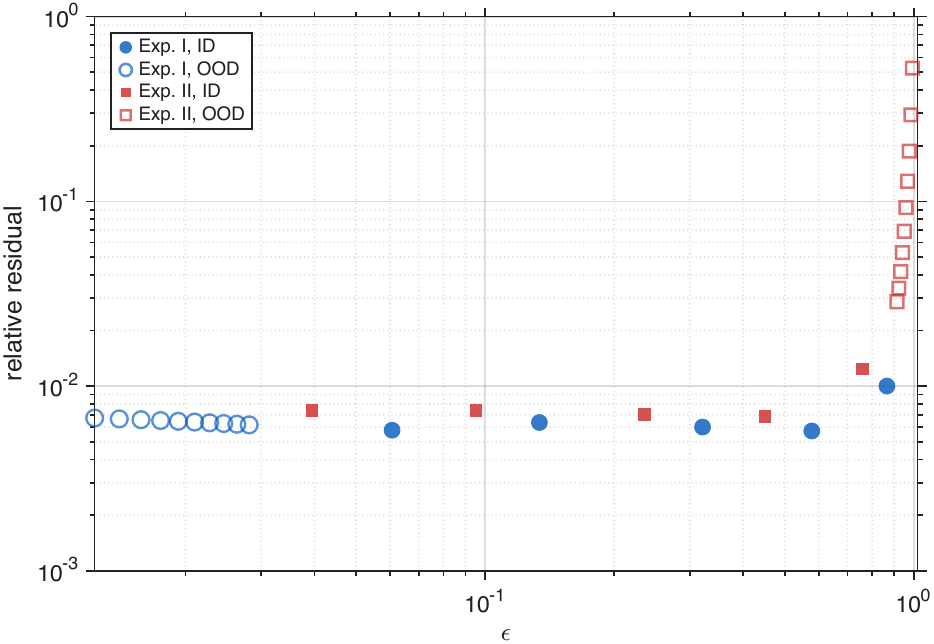}
\caption{Relative residuals of the predictions versus $\eps$ for both experiments. Filled and open markers denote ID and OOD solutions, respectively. The ID markers interleave because the experiments withhold different interior members. The OOD residuals of Experiment~I remain close to the ID level toward the free-fall limit, whereas those of Experiment~II increase sharply toward the linear limit.}
\label{fig:fnoresid}
\end{figure}

Table \ref{tab:resid} and Figure \ref{fig:fnoresid} report the relative residual, in \eqref{eq:relresid}, for both experiments. Reference solutions evaluated at the same temporal resolution give the baseline $\approx10^{-5}$ (see Section \ref{sec:criteria}). In Experiment~I, individual ID residuals range from $5.7\times10^{-3}$ to $1.0\times10^{-2}$, and OOD residuals increase monotonically from $6.17\times10^{-3}$ to $6.72\times10^{-3}$. In Experiment~II, OOD residuals increase monotonically from $2.86\times10^{-2}$ at $\eps=0.913$ to $5.26\times10^{-1}$ at $\eps=0.992$. Even for ID predictions, the residual lies two to three orders of magnitude above the baseline. Since the residual involves spatial and temporal derivatives, it is more sensitive to small-scale prediction errors than the $L^2$ measures in Sections \ref{sec:exp1} and \ref{sec:exp2}. 

\begin{table}[ht]
\centering
\caption{Average relative residuals of ID and OOD predictions for Experiment~I and Experiment~{II}. }
\label{tab:resid}
\begin{tabular}{lcc}
\toprule
 & ID & OOD \\
\midrule
Experiment I  & $6.10\times10^{-3}$ & $6.43\times10^{-3}$ \\
Experiment II & $7.45\times10^{-3}$ & $7.90\times10^{-2}$ \\
\bottomrule
\end{tabular}
\end{table}

The residual distinguishes the two extrapolation regimes without recourse to reference solutions. In Experiment~I, the average OOD residual is only $1.05$ times the ID residual, compared with a factor of $5.3$ for the $L^2$ error of $\eta$. Thus, the OOD predictions toward the free-fall limit lose accuracy near the crest at rest but remain consistent with the governing equations. 
In Experiment~II, on the other hand, the average OOD residual is $10.6$ times the ID value, and the individual residual increases monotonically toward the linear limit, from $2.86\times10^{-2}$ at $\eps=0.913$ to $5.26\times10^{-1}$ at $\eps=0.992$. As the wave amplitude vanishes, the absolute prediction error decreases too slowly to preserve the relative balance of \eqref{eq:resid}. This diagnostic therefore distinguishes a prediction that is inaccurate but dynamically plausible from one that has ceased to represent a standing wave, a distinction that the relative $L^2$ error alone does not make.

\section{Exploratory predictions beyond available data}\label{sec:epstry}

Near the free-fall limit $\eps\to0$, the acceleration of a fluid particle at the crest at rest approaches gravitational acceleration as the crest sharpens with little change in height, developing oscillations on progressively finer scales \cite{wilkening2011breakdown}. Indeed, the reference solutions computed in \cite{wilkening2012overdetermined} and used in the present study end at $\eps=0.0123$, where further numerical resolution becomes increasingly expensive. The objective of this section is then to examine the predictions from the two trained operators for candidate standing waves closer to free fall than any solution in the dataset. 

Note that these are exploratory predictions rather than OOD tests. No reference solutions are available at the chosen parameters, and assessments rely on the residual in \eqref{eq:relresid}, the predictions themselves, and comparisons between the two operators. This is the setting in which the residual serves as an independent diagnostic, as discussed in Section \ref{sec:criteria}. 

Since the operator takes the rest surface profile as input, not the parameter value, candidate rest profiles must first be constructed. We accomplish this by pointwise quadratic extrapolation. At each grid point $x_i$, in \eqref{eq:xgrid}, the rest surface elevation $\eta_0^\eps(x_i)$ of the ten steepest reference solutions, for $\eps\in[0.012,0.028]$, are fitted by a quadratic polynomial in $\eps$ and extrapolated to the targets $\eps\in\{0.010,0.008,0.006\}$. The resulting profiles continue the observed trend. Their crest heights increase slightly from $0.8251$ to $0.8263$, compared with $0.8245$ at $\eps=0.0123$, while their maximal slopes increase from $1.167$ to $1.296$, compared with $1.130$ for the reference endpoint. 
Additionally, we monitor whether the predicted surface elevation at $\tau=0$ reproduces the input profile, alongside the subsequent evolution.

\begin{figure}[htbp]
\centering
\includegraphics[width=\textwidth]{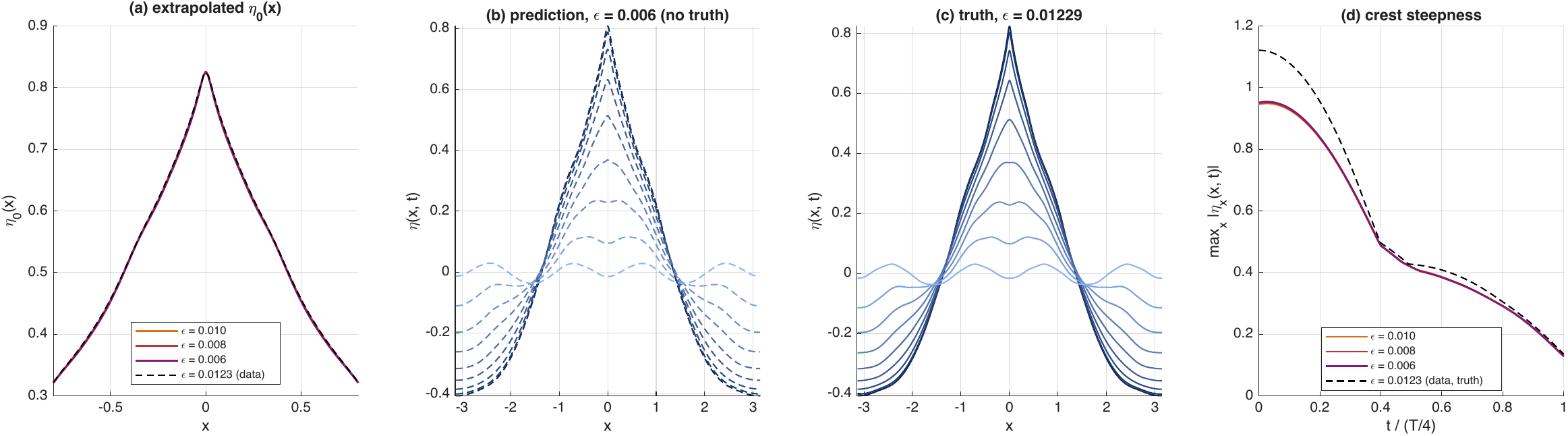}
\caption{Exploratory predictions from the Experiment~I operator beyond the available reference solutions. (a)~Extrapolated rest surface profiles. (b)~Predicted evolution at $\eps=0.006$, for which no reference solution is available. (c)~Reference evolution of the common benchmark at $\eps=0.0123$. (d)~Predicted maximal slope over the quarter period, compared with the benchmark steepness history.}
\label{fig:epstry1}
\end{figure}

\begin{figure}[htbp]
\centering
\includegraphics[width=\textwidth]{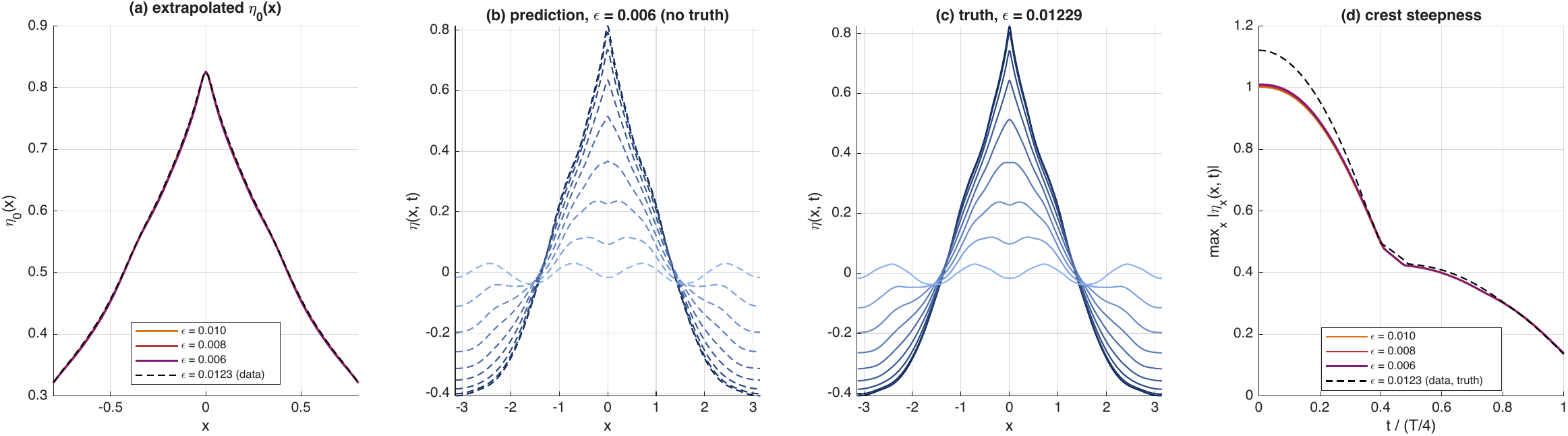}
\caption{Exploratory predictions from the Experiment~II operator using the same candidate inputs and panel layout as Figure \ref{fig:epstry1}. Panel (c) shows the identical reference benchmark at $\eps=0.0123$ to facilitate comparison between two operators.}
\label{fig:epstry2}
\end{figure}

Each of the two trained operators receives the same three candidate profiles, constructed solely from the reference data. Figures \ref{fig:epstry1} and \ref{fig:epstry2} show the results using a common four-panel layout: (a) the constructed rest surface profiles against the final member of the dataset, (b) the predicted evolution over the quarter period at the smallest target $\eps$, shown with dashed curves to indicate that no reference solution is available, (c) the reference evolution for the benchmark solution at $\eps=0.0123$, and (d) the predicted maximal slope $\max_x|\eta_x(x,t)|$ over the quarter period, compared with the reference steepness history. The benchmark in panel (c) is identical in both figures, the closest available reference member, regardless of which members are held out for an  experiment.

Panel (d) tests whether the predicted evolution is compatible with a smooth continuation of the observed family. The candidate profiles at rest are steeper than the benchmark by construction, so that such a continuation would begin above its maximal-slope curve and relax over the period in the same manner, without additional oscillations or crossings absent from the reference solution family. The two operators were trained on different portions of the branch. Disagreement between their predictions for identical inputs therefore provides an inexpensive proxy for model uncertainty. Agreement supports reproducibility across the two training partitions, but cannot exclude a shared bias.

We evaluate the residual without modification, but determining the time step $\Delta t=\frac{T/4}{M-1}$ requires the period of a candidate solution. We estimate the unknown period by quadratic extrapolation of $T^\eps$ over the same interval $\eps\in[0.012,0.028]$ used to construct the candidate rest profiles. The resulting residuals, reported in Table \ref{tab:residtry}, lie in the range $6.8$--$8.0\times10^{-3}$ for the two operators, comparable to the ID residuals and the Experiment~I OOD residual in Table \ref{tab:resid}. They change by less than one percent under percent-level perturbations of the assumed period. This diagnostic  detects no substantial deterioration relative to the validated range, but it does not establish that the predictions satisfy the governing equations or continue the solution branch.

\begin{table}[ht]
\centering
\caption{Relative residuals for exploratory predictions from the two operators. No reference solutions are available at the target parameter values. Residuals are evaluated using periods obtained by quadratic extrapolation over the same branch segment used to construct the rest surface profiles.}
\label{tab:residtry}
\begin{tabular}{lcc}
\toprule
$\eps$ (target label) & $r$ (Experiment I) & $r$ (Experiment II) \\
\midrule
$0.010$ & $6.81\times10^{-3}$ & $7.84\times10^{-3}$ \\
$0.008$ & $6.90\times10^{-3}$ & $7.89\times10^{-3}$ \\
$0.006$ & $7.00\times10^{-3}$ & $7.95\times10^{-3}$ \\
\bottomrule
\end{tabular}
\end{table}

Two limitations are central to the interpretation. First, pointwise quadratic extrapolation assumes a smooth continuation below $\eps=0.0123$. It cannot capture the increasingly fine oscillatory structure reported in \cite{wilkening2011breakdown}. Second, spectral truncation becomes more restrictive as the crest sharpens, potentially biasing the predictions toward smoother profiles. These results therefore assess a smooth continuation hypothesis with two independently trained surrogates. They do not constitute newly computed standing waves.

\section{Conclusions}\label{sec:conclusions}

We have formulated an operator learning problem for a one-parameter family of space-time periodic standing waves in infinite depth and approximated the solution operator using an FNO architecture. Given the rest wave profile and a normalized observation time, the learned operator returns the fluid surface, the velocity potential and dipole density at the surface at the time in one forward evaluation. This avoids time stepping and the nonlinear boundary value solve required for each reference solution. 
Trained on spectrally accurate reference solutions, the operator achieves comparable accuracy for all three fields within the training range, and the ID error has no systematic dependence on $\eps$ in Experiment~I.

The two OOD experiments reveal distinct sources of error. Toward the free-fall limit, the error increases monotonically as the crest sharpens, is concentrated near the crest and the instant of rest, and is largest for the fluid surface elevation. The steepest reference profiles retain appreciable Fourier content above the truncation wavenumber for the spectral convolution, whereas the ID profiles do not. This is consistent with a truncation-related loss of accuracy, although the dependence on the cutoff has not been quantified. Toward the linear limit, the absolute error does not increase while the wave amplitude decreases in proportion to $1-\eps$ as $\eps \to 1$. The resulting increase in relative error is consistent with the standardization of the fields by constants fixed on the training set, which does not give additional weight to discrepancies that are small in absolute terms but significant for small amplitude waves.

The residual of the governing equations, evaluated from the predictions, provides a complementary assessment without reference solutions. ID residuals lie two to three orders of magnitude above the reference baseline, illustrating the sensitivity of derivative-based diagnostics to small prediction errors. Toward the free-fall limit, the OOD residual remains close to the ID level, and the predictions lose accuracy near the crest without becoming less consistent with the governing equations. Toward the linear limit, the residual increases monotonically and for the member nearest the limit, reaches approximately half the root-mean-square magnitude of the dominant linear terms. This indicates a substantial violation of the tested dynamic balance.  

For exploratory inputs beyond the final reference member, both operators produce residuals close to the ID level when evaluated with extrapolated periods. These inputs are constructed by smooth extrapolation, which cannot represent the increasingly fine-scale branch structure near free fall \cite{wilkening2011breakdown}. The predictions must consequently be interpreted as an assessment of a smooth continuation hypothesis, rather than as new standing wave solutions.

Several extensions follow. Predictions within and outside the available data could provide initial guesses for the overdetermined shooting method of \cite{wilkening2012overdetermined}, allowing a conventional solver to refine and verify candidate solutions. Larger spectral cutoffs, crest-adapted coordinates, and multi-resolution architectures warrant investigation near free fall. Amplitude-adapted normalization or a loss based on the relative error of each solution may improve accuracy near the linear limit. In a physics-informed neural operator formulation \cite{li2024physics}, the present residual could supplement the training loss, with its limited scope explicitly recognized, and the period could be included as an additional output. Finally, extensions to finite depth, three dimensions, and capillary-gravity waves with the effects of surface tension could be combined with reduced models based on proper orthogonal decomposition \cite{williams2012low}. 

\paragraph{Acknowledgements}
The authors gratefully acknowledge Jon Wilkening of the University of California, Berkeley for generously providing the essential numerical data used in this study. Access to this high-resolution dataset was crucial for guiding the direction of our research and circumvent the need for computationally intensive reference computations.

SA was supported  by the NSF Graduate Research Fellowship Program under DGE 21-46756, the Alfred P. Sloan Foundation, and the Sloan University Center of Exemplary Mentoring at Illinois when this project began, when she was a PhD student at the University of Illinois Urbana-Champaign.
VMH is partially supported by the National Science Foundation under DMS-2475358. DQ is partially supported by Office of Naval Research under N00014-24-1-2192 and National Science Foundation under DMS-2407361. 

\paragraph{Author Contributions}
CM performed the primary numerical computations, analyzed the production data, and wrote the original draft of the manuscript. SA conducted the preliminary numerical investigations that motivated the study and contributed to the writing and critical review of the initial manuscript. DQ provided technical expertise and methodological guidance, and supervised the computational framework. VMH conceptualized the project, secured resources, coordinated the research team, and revised and edited the final manuscript. All authors reviewed and approved the final manuscript.

\paragraph{Conflict of Interest}
The authors declare no competing financial or personal interests.

\paragraph{Data Availability}
The data are available upon reasonable request.
The code is available at \url{https://github.com/chhmou/fno-sw}.
\bibliographystyle{unsrtnat}
\bibliography{ref}

\end{document}